\documentclass[aps,preprint,nofootinbib,showpacs]{revtex4}
\usepackage{graphicx,color,amsmath, epsfig}
\newcommand{\be}{\begin{equation}}
\newcommand{\ee}{\end{equation}}
\newcommand{\ba}{\begin{eqnarray}}
\newcommand{\beq}{\begin{equation}}
\newcommand{\eeq}{\end{equation}}
\newcommand{\ea}{\end{eqnarray}}
\def\bea{\begin{eqnarray}}
\def\eea{\end{eqnarray}}

\def\err#1#2{\lower2pt\hbox{ $\stackrel{\scriptstyle +#1}{\scriptstyle -#2}$}}
\def\ga{\mathrel{\raise.3ex\hbox{$>$\kern-.75em\lower1ex\hbox{$\sim$}}}}
\def\la{\mathrel{\raise.3ex\hbox{$<$\kern-.75em\lower1ex\hbox{$\sim$}}}}

\begin{document}

\preprint{%
\vbox{%
\hbox{December'07}
}}
\title{Charged Higgs bosons in the Next-to MSSM (NMSSM)}

\author{A.G. Akeroyd$^{1,2}$, Abdesslam Arhrib$^3$, Qi-Shu Yan$^{4}$}
\affiliation{$^1$ Department of Physics, National Cheng Kung University, 
Tainan 701, Taiwan} 
\affiliation{$^2$ National Center for Theoretical Sciences, Taiwan}
\affiliation{$^3$ Department of Physics, National Central University, 
Chung Li, Taiwan \\
and D\'epartement de Math\'ematiques, Facult\'e des Sciences et 
Techniques, B.P 416 Tangier, Morocco}
\affiliation{$4$ Department of Physics, National Tsing Hua University, 
Hsinchu, Taiwan}

\date{\today}
\begin{abstract}
The charged Higgs boson decays 
$H^\pm\to W^\pm A_1$ and $H^\pm\to W^\pm h_i$ are studied in the 
framework of the next-to Minimal Supersymmetric Standard Model (NMSSM).
It is found that 
the decay rate for $H^\pm\to W^\pm A_1$ can exceed the rates for 
the $\tau^\pm\nu$ and $tb$ channels both below and above the 
top-bottom threshold. The dominance of $H^\pm\to W^\pm A_1$
is most readily achieved when $A_1$ has a large doublet component and
small mass. We also study the production process $pp\to H^\pm A_1$ 
at the LHC followed by the decay $H^\pm\to W^\pm A_1$ 
which leads to the signature $W^\pm A_1 A_1$.  We suggest that  
$p p\to H^\pm A_1$ is a promising discovery channel for a
light charged Higgs boson in the NMSSM with small or moderate $\tan\beta$ 
and dominant decay mode $H^\pm \to W^\pm A_1$.
This $W^\pm A_1 A_1$ signature can also arise from the Higgsstrahlung process
$pp\to W^\pm h_1$ followed by the decay $h_1\to A_1 A_1$.
It is shown that there exist regions of parameter space 
where these processes can have comparable cross sections and
we suggest that
their respective signals can be distinguished at the LHC by
using appropriate reconstruction methods. 
\end{abstract}
\maketitle
 
\section{Introduction}
An attractive extension of the Minimal Supersymmetric
Standard Model (MSSM) is the Next-to MSSM
(NMSSM) in which an additional singlet neutral complex scalar
field $S$ is added. The presence of this singlet field 
provides an elegant solution to the $\mu$ problem of the MSSM.
The $\mu$ parameter in the MSSM superpotential,
which does not break supersymmetry (SUSY) and is present when SUSY is unbroken,
is completely unrelated to the electroweak or SUSY breaking scales.
In some models like Supergravity, 
$\mu$ is naturally expected to be of the order $M_{\rm{Planck}}$. 
However, the radiative electroweak symmetry breaking conditions 
require the $\mu$ parameter to be of the same order as $M_Z$. 
Such a conflict is called the $\mu$ problem \cite{mu}.

The superpotential of the NMSSM contains the term 
$\lambda \hat{S} \, \hat{H}_u \, \hat{H}_d$, and
the $\mu$ term of the MSSM which mixes the two doublet 
fields $\hat{H}_u$ and $\hat{H}_d$ is not present explicitly.
When the singlet field acquires a vacuum expectation
value $<s>$ of the order of the SUSY breaking scale,
an effective $\mu$ parameter $\mu_{eff}=\lambda s$ of the order of 
the electroweak scale is then dynamically generated.
Moreover, it has been shown that with the additional singlet Higgs field
the MSSM fine-tuning (or ``little hierarchy problem'') problem 
can be ameliorated in regions of the NMSSM parameter space
\cite{Dermisek:2005ar,Dermisek:2007yt}.

A charged Higgs boson ($H^\pm$) appears in any extension of the
Standard Model with two hypercharge Y=1 doublets. Its phenomenology has
been extensively studied in both the Two Higgs Doublet Model
(2HDM) and MSSM. The phenomenology of $H^\pm$ in the NMSSM
is similar in many ways to that in the MSSM since no 
charged singlet fields have been added. 
The increased parameter content of the
NMSSM scalar potential compared to that of the MSSM permits
large mass splittings among the Higgs spectrum, which allows
other decay modes of $H^\pm$ to be important which were substantially 
suppressed in the context of the MSSM.
In the MSSM the coupling $H^\pm A W$ (where $A$ is the
CP-odd neutral Higgs boson) contains
no mixing angle suppression but the relation $M_A\sim M_{H^\pm}$ 
ensures that the decay $H^{\pm} \to A W$ is greatly suppressed
in most of the parameter space \cite{Moretti:1994ds},\cite{Djouadi:1995gv}.
In the NMSSM there are two pseudoscalars $A_1$ and $A_2$ 
which are mixtures of the doublet and singlet fields.
There exists regions in the theoretical parameter space where
$A_1$ is predominantly doublet and light, and hence 
the decay $H^{\pm} \to A_1 W$ is unsuppressed.
 
The importance of the decay $H^{\pm} \to A_1W$ in the NMSSM 
was emphasized in
\cite{Drees:1998pw} where it was shown that
dominance over $H^\pm\to cs,\tau\nu$ is possible and  
branching ratios close to 100\% can be attained 
for intermediate values of $\tan\beta$.
A LHC simulation was performed in \cite{Drees:1999sb} and concluded
that such a decay offers very good detection prospects 
for $H^\pm$
if the branching ratios of $t\to H^\pm b$ and $H^\pm \to A_1W$ 
are sufficiently large. In this work we perform a comprehensive 
scan of the NMSSM parameter space using the publicly available code
NMHDECAY \cite{Ellwanger:2004xm}
in order to identify the regions
where $H^\pm \to A_1W$ can be sizeable.
 
The strength of the coupling $H^\pm A_1W$ can also have an
application to the production of $H^\pm$ via
$pp \to H^\pm A_1$ which has been studied in the CP conserving 
MSSM \cite{Kanemura:2001hz},\cite{Belyaev:2006rf} and CP
violating MSSM \cite{Akeroyd:2003jp}. 
If the branching ratio for the 
decay $H^\pm \to A_1W$ were also sizeable such a production mechanism
would lead a final state of $Wbbbb$ (for $M_{A_1}>2m_b$) 
\cite{Akeroyd:2003jp} which 
has been simulated \cite{Ghosh:2004wr} in the context of the LHC with 
promising conclusions.
This $Wbbbb$ signature can also arise from
the process $pp \to Wh_1\to W A_1A_1$ which was simulated in 
\cite{Cheung:2007sv}
and shown to provide a clear signal at the LHC. 
We compare the magnitude of both
mechanisms and discuss how they may be distinguished.

Our work is organized as follows: in section II we present a short review
of the Higgs sector of the NMSSM; in section III 
the limits that lead to a light $A_1$ in the NMSSM parameter space 
are listed;
in section IV the phenomenology of $H^\pm$ is introduced;
section V contains our numerical results for the branching ratios
of $H^\pm \to A_1 W,h_1 W$ and cross-sections 
$pp \to H^\pm A_1\to Wbbbb (W\tau\tau\tau\tau)$
and $pp \to W h_1\to Wbbbb (W\tau\tau\tau\tau)$.
Conclusions are given in section VI.

\section{A brief review on the Higgs sector of 
the NMSSM}
For detailed discussions of the Higgs sector of the 
NMSSM the reader is referred to \cite{Drees:1988fc,Elliott:1993bs,
Franke:1995tc,Miller:2003ay,Accomando:2006ga}.
In this section we follow the notation of Ref.~\cite{Ellwanger:2004gz}.
The NMSSM Higgs sector differs from that of the MSSM by the
addition of an extra complex scalar field, $S$. The Higgs fields of
the model then consist of the usual two Higgs doublets $\hat{H}_u$ and
 $\hat{H}_d$ together with this extra Higgs singlet.

In the NMSSM Lagrangian, the extra singlet field is allowed to couple
only to the Higgs doublets of the model and consequently the
couplings of the new field $S$ to gauge bosons and fermions
will only be manifest via their mixing with the doublet Higgs fields. 
The superpotential of the NMSSM is given by
\begin{equation}
\label{eq:superpotential}
W=W_{\rm{MSSM}} + \lambda \hat{S} \, 
\hat{H}_u \, \hat{H}_d+ \frac{1}{3}\kappa \, \hat{S}^3.
\end{equation}
where $W_{\rm{MSSM}}$ is the usual MSSM superpotential and
only terms that depend on the singlet field are explicitly written.
The soft breaking terms for both the doublet and singlet
are included in $V_{\rm soft}$:
\begin{eqnarray}
V_{\rm soft}&=&m_{H_u}^2|H_u|^2 + m_{H_d}^2|H_d|^2 + m_S^2|S|^2
+ [\lambda A_{\lambda}SH_uH_d+\frac{1}{3}\kappa A_{\kappa}S^3+\textrm{h.c.} ]
\;, \label{eq:HpotS}
\end{eqnarray}
The parameters additional to those of the MSSM are:
$\lambda$, $\kappa$, $A_{\lambda}$, $A_{\kappa}$, $m_S$ and
the vacuum expectation value of the singlet field, $s$,
which will generate the effective $\mu$ term given
by $\mu_{eff}=\lambda s $. 
As in the MSSM, $m_S$ can be fixed by the minimization condition
of the scalar potential.

After electroweak symmetry breaking
the Higgs spectrum of the NMSSM consists of three neutral 
scalars ($h_1,h_2,h_3$), two pseudoscalars ($A_1,A_2$) 
and a pair of charged Higgs bosons $H^\pm$. In both the CP-odd
and CP-even sector the physical eigenstates are ordered as
$M_{h_1}\la M_{h_2} \la M_{h_3}$ and $M_{A_1}\la M_{A_2}$.
The mass of $H^\pm$ at tree-level is given by \cite{Drees:1988fc},
\cite{King:1995ys}:
\begin{equation}
M_{H^\pm}^2 =\frac{2 \mu_{eff}}{\sin 2\beta} (A_\lambda +\kappa s)+M_W^2
-\lambda^2 v^2 \label{chargedhiggs}
\end{equation}
where $\tan \beta =v_u/v_d$  and $v^2=v_u^2+v_d^2$.
This differs from the corresponding MSSM expression in which 
$M_A$ and $M_{H^\pm}$ are strongly correlated and become 
roughly equal for $M_A\geq 140$ GeV. 

The CP-odd mass matrix can be obtained as follows:
Firstly, as in MSSM one rotates the bare 
fields $(\Im\rm{m}H_u, \Im\rm{m}H_d,\Im\rm{m}S)$
into a basis $(A,G,\Im\rm{m}S)$ where $G$ is a massless Goldstone
boson.
Then one eliminates the Goldstone mode and the remaining 
$2\times 2$ CP-odd mass matrix in the basis $(A,\Im\rm{m}S)$ 
is given by:
\begin{eqnarray}
{\cal M}_{P,11}^2 & = &  \frac{\lambda s }{\sin\beta \cos\beta}\, 
(A_\lambda +  \kappa s ), \nonumber\\
{\cal M}_{P,22}^2 & = & ( 2 \lambda \kappa + 
\frac{\lambda  A_\lambda }{ 2 s}) \sin 2\beta v^2\,
-3 \kappa A_\kappa s, \nonumber\\
{\cal M}_{P,12}^2 & = & \lambda v \, 
(A_\lambda - 2\kappa s).
\end{eqnarray}

Here $A=\cos\beta \Im\rm{m}H_u + \sin\beta \Im\rm{m}H_d$ 
is the CP-odd MSSM Higgs boson while $\Im\rm{m}S$ comes 
from the singlet $S$ field. 
The pseudoscalars fields are further rotated to the diagonal basis
($A_1$, $A_2$) by an orthogonal $2\times 2$ matrix such that:

\begin{eqnarray}
A_1&=& \cos\theta_A A + \sin\theta_A \Im\rm{m}(S) \nonumber\\
A_2&=& -\sin\theta_1  A + \cos\theta_A \Im\rm{m}(S) 
\end{eqnarray}
where
\begin{eqnarray}
\cos\theta_A &=&
\frac{ {\cal M}_{P,12}^2 }{\sqrt{{\cal M}_{P,12}^4 + 
(M_{A_1}^2-{\cal M}_{P,11}^2)^2 }}\quad , \quad 
\sin\theta_A =\frac{ M_{A_1}^2 -  {\cal M}_{P,11}^2  }{\sqrt{{\cal M}_{P,12}^4 + 
(M_{A_1}^2-{\cal M}_{P,11}^2)^2}}
\label{cpmixing}
\end{eqnarray}

The Higgs boson-gauge boson couplings originate from the covariant derivative
of the kinetic energy term. Those relevant for our study
are described by the following Lagrangian:
\begin{eqnarray}
{\cal L}_{VVH,VHH}& =&
gm_W g_{VVh_i}  W^{+\mu} W_{\mu}^- h_i -
g W_\mu^+(\frac{ig_{W^+H^-h_i}}{2} h_i  
+\frac{P_{i1}}{2} A_i)\stackrel{\leftrightarrow}{\partial}^\mu H^-
+ h.c
\label{gaugecoup}
\end{eqnarray}
where $g_{VVh_i}=\sin\beta S_{i1} +\cos\beta S_{i2}$, 
$g_{W^+H^-h_i}= \cos\beta S_{i1}-\sin\beta S_{i2}$, $P_{11}=\cos\theta_A$ 
and $P_{21}=-\sin\theta_A$, $S$ and $P$ are orthogonal matrix which 
diagonalize 
respectively the CP-even and CP-odd scalar mass matrix.
>From the last term in eq.~(\ref{gaugecoup}) one can see that the vertex
 $W^\pm H^\mp A_1$ is directly proportional to $P_{11}$
i.e. the doublet component of the mass eigenstate $A_1$. 
Consequently, if $A_1$ is entirely composed of doublet fields   
this coupling is maximized and if $A_1$ is purely singlet
the coupling vanishes.

As in the MSSM one can easily derive the following sum rules:
\begin{eqnarray}
&& \sum_{i=1}^3 g_{WWh_i}^2 =1\nonumber\\
&&  g_{WWh_i}^2 + g_{W^+H^-h_i}^2 + S_{i3}^2=1 \quad i=1,2,3
\label{sumrule}
\end{eqnarray}
Here $S_{i3}$ is the singlet component of $h_i$. From the second sum rule
it follows that if $h_i$ is purely doublet ($S_{i3}\approx 0$) then 
the MSSM sum rule, $g_{WWh_i}^2 + g_{W^+H^-h_i}^2=1$, 
is recovered where $h_i$ is entirely composed of doublet
fields. Conversely, if $h_i$ is purely singlet ($S_{i3}^2\approx 1$) then
one has $g_{WWh_i}^2 + g_{W^+H^-h_i}^2\approx 0$ and
both $h_iVV$ and $h_iH^+W^-$ must be suppressed, and this will present a
real challenge for the detection of Higgs bosons. This sum rule will be 
explored in our numerical analysis.

\section{A light $A_1$ in the NMSSM parameter space}

The parameter space of the NMSSM can naturally accommodate 
a light $A_1$ which is of great phenomenological interest.
To identify such regions it is instructive to examine the vanishing limits of 
the determinant of the mass matrix of the pseudoscalar, 
which can be expressed as:
\bea
Det M^2_P &=& - \frac{3 \kappa \lambda s}{\sin 2 \beta} \Bigg (
2 \kappa s^2 A_\kappa + 2 s A_\kappa A_\lambda - 3 \lambda  
A_\lambda v^2 \sin 2 \beta \Bigg ) \,.
\eea
It is then straightforward to identify four distinct cases where 
$Det M^2_{P}$ approaches $0$:
\begin{itemize}
\item Case 1: $A_{\lambda} \rightarrow 0 $
and $A_{\kappa} \rightarrow 0$ \cite{NMSSMR},
\item Case 2: $\kappa \to 0$ \cite{NMSSMPQ},
\item Case 3: $\lambda \to 0$,
\item Case 4: $s \to 0$.
\end{itemize}
Moreover, it is evident that combinations of these basic cases can 
also lead to a light $A_1$. The requirement of 
perturbativity up to the grand unification scale restricts 
$\lambda< 0.8$ \cite{ellis}. 
Therefore Case 4 ($s\to 0$) is 
ruled out since it would lead to a very small $\mu_{eff}$ which 
is excluded by the mass bound for charginos from direct searches.
However, if one gives up this perturbative requirement
up to grand unification scale and considers $\lambda\gg 1$, 
as in the so-called $\lambda$SUSY model \cite{Cavicchia:2007dp}
(which can be realized in the supersymmetric fat Higgs models \cite{fat}),
then Case 4 might be viable.

The first two limits are related with
the discrete symmetries of Higgs potential: 
one is called the R-axion limit with $A_{\lambda} \rightarrow 0 $
and $A_{\kappa} \rightarrow 0$ \cite{NMSSMR};
the other is called the PQ-axion limit with $\kappa \to 0$
the superpotential eq.(\ref{eq:superpotential}) and 
its associated Lagrangian contains an extra 
global $U(1)$ symmetry \cite{NMSSMPQ}. 
In both cases, these symmetries are spontaneously broken
by the Higgs vev leading to Pseudo-Goldstone boson in the spectrum.

At tree-level, in the R-axion limit \cite{NMSSMR}, the mass 
spectra and mixing of the CP-odd Higgs sector can be
expressed as: 
\bea
m_{A_1}^2 &=& 3 s (- \kappa A_{\kappa} \sin^2 \theta_A +  
\frac{3}{2 \sin 2 \beta } \lambda A_{\lambda} \cos^2 \theta_A ) 
+ O(\kappa^2 A_{\kappa}^2, \lambda^2 A_{\lambda}^2)\,,\nonumber\\
m_{A_2}^2 &=& \frac{2 \lambda \kappa v^2}{\cos^2 \theta_A} \sin 2 \beta
+ O(\kappa^2 A_{\kappa}^2, \lambda^2 A_{\lambda}^2)\,,\nonumber\\
\tan \theta_A &=& \frac{s}{v \sin 2 \beta} 
+ O(\kappa A_{\kappa}, \lambda A_{\lambda})\,.
\label{raxion}
\eea
In the R-axion limit scenario, as can be seen from eq.(\ref{raxion}),
a light pseudoscalar is obtained for small
$\kappa A_{\kappa}$ and $\lambda A_{\lambda}$ or  
a combination of small $\kappa A_{\kappa}$ and  $\lambda A_{\lambda}$.

At tree-level, in the PQ-axion limit \cite{NMSSMPQ}, one has:
\bea
m_{A_1}^2 &=&  3 s \kappa (- A_{\kappa} \sin^2 \theta_A + 
\frac{6} {\sin 2 \beta} \lambda s \cos^2 \theta_A  )
+ O(\kappa^2)\,,\nonumber\\
m_{A_2}^2 &=& - \frac{2 \lambda A_{\lambda} v}{\sin 2 \theta_A} + 
O(\kappa^2)\,,\nonumber\\
\tan \theta_A &=& - \frac{2 s}{v \sin 2 \beta} 
+ O(\kappa)\,.
\label{PQaxion}
\eea
It is interesting to see that in eq.~(\ref{PQaxion})
 the limit $\kappa\to 0$ gives $m_{A_1}\to 0$.
This is actually the case where the U(1) PQ symmetry is left unbroken 
in the superpotential. The spontaneous breaking of such PQ symmetry
by a Higgs vev leads to a massless Goldstone boson, the axion.
To obtain a light pseudoscalar $A_1$ one needs to introduce
a small $\kappa$ which only slightly breaks the PQ symmetry.

The third case is also related with a 
discrete symmetry of two Higgs doublet models. In this limit one has:
\bea
m_{A_1}^2 &=& \frac{ 2 \lambda s}{\sin 2 \beta} (A_\lambda + \kappa s)
+ O(\lambda^2)\,,\nonumber\\
m_{A_2}^2 &=& - 3 \kappa s A_\kappa + O(\lambda)\,,\nonumber\\
\tan \theta_A &=& \lambda \frac{(A_\lambda - 2 \kappa s) v}{3 
\kappa A_\kappa s} + O(\lambda^2)\,.
\label{case3}
\eea
When $\lambda \to 0$, a large value for $s$ is needed to 
keep $\mu_{eff}$  of the order of the electroweak scale.
In this case $\lambda \to 0$, and for $\mu_{eff}$ fixed,   
$A_1$ is mainly doublet and this is the exact MSSM limit.

\section{$H^\pm$ in the NMSSM}
In this section we describe the phenomenology of the $H^\pm$
in the NMSSM and highlight its differences with the phenomenology
of $H^\pm$ in the MSSM.
The phenomenology of $H^\pm$ in the NMSSM 
has many similarities with that of $H^\pm$ in the MSSM 
(the latter recently reviewed in \cite{Roy:2004az}).
This is to be expected since the fermionic couplings 
are identical in the two models.
The main differences in their phenomenology originate from the possibility 
of large mass splittings among the Higgs bosons in the NMSSM which permits 
decay channels like $H^{\pm} \to A_1W$ to proceed on-shell \cite{Drees:1998pw}. 
In the MSSM such a decay can only be open for extreme choices of
certain SUSY parameters (e.g. for $\mu> 4M_{SUSY}$ \cite{Akeroyd:2001in})
which induce large quantum corrections in the effective scalar
potential.  Moreover, in the NMSSM a light CP-even $h_1$ is also allowed and 
one can have the opening of the  decay $H^{\pm} \to h_1W$ both below and 
above the top-bottom threshold. This latter channel may change the NMSSM 
phenomenological predictions for
the charged Higgs with respect to the MSSM \cite{Drees:1998pw}.
In the MSSM the decay  $H^{\pm} \to h_1W$ is also open 
but the coupling $g_{W^+H^-h_1}\sim \cos^2(\beta-\alpha)$ is strongly
suppressed when $M_{H^\pm}\gg m_{h_1}+m_W$ and thus its branching
ratio is very small for such $M_{H^\pm}$.  For 
$M_{H^\pm}< m_{h_1}+m_W$ and just above the threshold the branching 
ratio for this channel can reach $10\%$ at most for small values 
of $\tan\beta$ \cite{Moretti:1994ds}, \cite{Djouadi:1995gv}, 
\cite{Drees:1999sb}.
 
The phenomenology of $H^\pm$ in the NMSSM has received considerably
less attention than its neutral Higgs sector. In recent
years much effort has been focused on establishing a "no--lose
theorem'' at the LHC in which detection of at least one Higgs boson
in the NMSSM is guaranteed. However, the potential 
importance of the decay $h_1 \to A_1A_1$ 
\cite{NMSSMR},\cite{Gunion:1996fb} has 
prevented such a theorem being established \cite{Ellwanger:2004gz},
\cite{Ellwanger:2001iw}, \cite{Ellwanger:2003jt},
\cite{Ellwanger:2005uu}, \cite{Moretti:2006hq}.
Moreover, it has been shown that a large branching ratio for $h_1 \to A_1A_1$
would weaken the LEP bounds for a SM like $h_1$ in the NMSSM
\cite{Dermisek:2005ar}. 

For $M_{A_1}< 2m_b$ \cite{Dermisek:2005gg,Dermisek:2007yt}
dominance of $h_1\to A_1A_1$ has the virtue of allowing 
$h_1$ as light as $90\to 100$ GeV and can realize the
"LEP excess scenario" easily. Such values of $M_{h_1}$ can be accommodated in the NMSSM with little fine-tuning, in contrast to the MSSM 
case where considerable fine-tuning is necessary in order to
comply with the LEP limit $M_{h_1}> 114$ GeV from the Higgsstrahlung
channel. However, a large branching ratio for $h_1\to A_1A_1$
followed by $A_1A_1\to 4\tau$ is challenging for detection at the
Tevatron (see \cite{Graham:2006tr}).
For the final states $V2b2\tau$ and $V4b$ at the Tevatron, the observation is difficult due to the limited statistics, as shown in \cite{Carena:2007jk}. At the LHC, by utilizing the central exclusive production process and high-resolution low-angle sub-detectors, it is shown in 
\cite{Forshaw:2007ra} that it is possible to reconstruct the masses of $h_1$ and $A_1$. In \cite{Arhrib:2006sx} it was suggested that production of 
$A_1$ in association with charginos followed by
the possibly dominant decay 
$A_1\to \gamma\gamma$ could offer good detection prospects
for an almost purely singlet $A_1$. An alternative probe is
the decay $\Upsilon\to A_1\gamma$ at B factories
\cite{Dermisek:2006py}. A high-energy $e^+e^-$ 
linear collider would easily probe the scenario of dominant decay 
$h_1\to A_1A_1$ for $m_{A_1} < 2m_b$ via the recoil mass technique 
which is insensitive to the decay of $h_1$.

For $M_{A_1}> 2m_b$ one would have the dominant decay
 $A_1A_1\to bbbb$ for which a LEP limit of $M_{h_1}> 110$ GeV was derived. In such a scenario the fine-tuning problem is not greatly ameliorated but detection prospects at the LHC are much better.
In partonic level analyses it has been shown that a signal with high significance and 
full Higgs mass reconstruction can be obtained from the process $pp \to Wh_1 \to WA_1A_1\to Wbbbb$ \cite{Cheung:2007sv,Carena:2007jk}. The main challenge in reconstructing 
the full decay chain is to retain an adequate tagging efficiency of b's in the low 
$p_T$ region where signal events are located, 
as shown in \cite{Carena:2007jk}.

In many of the studies which are concerned with establishing a no-lose 
theorem the charged Higgs mass is taken to be very heavy $M_{H^\pm} > 400$ GeV
(e.g. the benchmark points in \cite{Ellwanger:2005uu}).
It has been known for some time that
a moderately light $M_{H^\pm}<m_t$ is possible in the NMSSM. A first
detailed study appeared in \cite{Drees:1998pw}, and 
this possibility has recently been emphasized in \cite{Godbole:2007jq}.
However, such a $H^\pm$ would contribute sizably to the rare decay 
$b\to s\gamma$ whose branching ratio has been measured and 
is consistent with the SM expectation. In the context of the NMSSM
a contribution to $b\to s\gamma$ from another New Physics particle
(usually the lightest chargino, $\chi^\pm_1$) is needed
to partially cancel the large $H^\pm$ contribution for $M_{H^\pm}<m_t$
\cite{Bertolini:1990if}.
If flavour violation induced by gluinos ($\tilde g$) is considered
\cite{Borzumati:1999qt}, 
the NMSSM parameter space for a light $H^\pm$ can be enlarged 
while keeping the branching ratio for $b\to s\gamma$ consistent 
with the measured value.
This merely requires a suitable cancellation among the contributions
from $H^\pm$, $\chi^\pm_1$ and $\tilde g$, the latter being
of essentially arbitrary magnitude. In light of this possibility 
we do not impose the $b\to s\gamma$ constraint in our numerical
analysis. 

Another potentially important constraint on the
scenario of $M_{H^\pm}\la m_t$ comes from 
the measurement of the decay $B^\pm \to\tau^\pm\nu$  \cite{Ikado:2006un}.
This decay is mediated at tree-level \cite{Hou:1992sy} by $H^\pm$
and its contribution cannot be canceled by any other new particle 
in the model.
Current data excludes two regions in the parameter space of
$[M_{H^\pm},\tan\beta]$. However, the non-holomorphic contribution
\cite{Akeroyd:2003zr} would shift the location of these 
two regions and thus we do not impose such a constraint in our analysis.
Importantly, for $\tan\beta \la 20$ of most interest to us 
$M_{H^\pm}\la m_t$ is almost always allowed. 
Moreover, a recent analysis \cite{Domingo:2007dx} shows that
a light charged Higgs boson in the NMSSM is compatible 
with the constraints from $b\to s\gamma$, $\Delta M_q$, $B^\pm \to\tau^\pm\nu$
and $B_s\to \mu^+\mu^-$ even without invoking extra 
sources of flavour violation from gluinos.

\section{Charged Higgs decay $H^\pm \to S W$  and
         the production mechanism $pp\to H^\pm S$,  
$S=A_1,h_1$}
\subsection{Charged Higgs decay modes $H^\pm \to S W$}
The decay $H^\pm \to A W$, where $A$ is a CP-odd Higgs boson,
may be sizeable in a variety of models with a non-minimal Higgs sector
such as Two Higgs doublet models (Type I and II) 
\cite{Borzumati:1998xr,Akeroyd:1998dt,Akeroyd:2000xa} and in 
SUSY models with Higgs triplets \cite{DiazCruz:2007tf}.
Two LEP collaborations (OPAL and DELPHI) performed a search for 
a charged Higgs decaying to $A W^*$ (assuming $m_A> 2m_b$)
and derived limits on the charged Higgs mass
\cite{Abdallah:2003wd} comparable to those obtained from the
search for $H^\pm\to cs,\tau\nu$. In the MSSM the decay width 
for $H^\pm \to A W$ is very suppressed
in most of the parameter space \cite{Moretti:1994ds,Djouadi:1995gv}
because the charged Higgs and the CP-odd Higgs are close to mass
degeneracy.

The importance of the decays $H^\pm \to A_1 W$ and 
$H^\pm \to h_1 W$ in the NMSSM was first pointed out in \cite{Drees:1998pw}. 
Their branching ratios may be close to $100\%$
which can provide a clear signal at the LHC. Simulations of the process
$pp \to t\overline t$ followed by $t\to H^\pm b$  and  $H^\pm\to A_1W$
have been performed for the NMSSM \cite{Drees:1999sb}, CP conserving MSSM
\cite{Moretti:2000yg} and CP violating MSSM  \cite{Ghosh:2004cc}. 
The partial width is given by:
\begin{equation}
\Gamma(H^\pm \to A_1 W)=
\frac{\alpha \cos^2\theta_A}{16 s_W^2 M_W^2 M_{H^{\pm}}^3 } 
\lambda^{\frac{3}{2}}(M_{H^{\pm}}^2, M_{A_1}^2,M_W^2 ) \label{width0}
\end{equation}
where $\lambda(x,y,z)=x^2+y^2+z^2 -2(xy+ xz+ yz)$ 
is the two--body phase space function. The decay width of $H^\pm \to h_1 W$
can be obtained from eq.~(\ref{width0}) by replacing $\cos\theta_A$ by 
$g_{H^\pm W^\mp h_1}$ and $M_{A_1}$ by $M_{h_1}$.

As can be seen from (\ref{width0}), the decay width of $H^\pm \to A_1 W$
is directly proportional to $\cos\theta_A$ which is the doublet component
of $A_1$. This decay width can be substantially enhanced if 
$A_1$ is predominantly composed of doublet fields.
However, even with small doublet (large singlet) component of $A_1$ 
it is possible that $H^\pm \to A_1 W$ is the dominant decay mode. 
We perform a scan of the parameter space using the code 
 \cite{Ellwanger:2004xm}  (NMSSM-Tools incorporates the 
LEP2 bounds for 4b and 6b final states) in order to quantify the importance
of $H^\pm \to A_1 W$ and $H^\pm \to h_1 W$.

Hereafter we assume that all scalar 
superparticles share the same soft mass term
$M_{SUSY}$, and the ratios of gaugino masses satisfy $M_1 : M_2 : M_3
= 1:2:6$; the trilinear couplings are related to $M_{SUSY}$ but
the sign is not fixed, {\it i.e.} $A_{t,b} = \pm 2 M_{SUSY}$. 
We scan the parameter space of the model by 
varying the free parameters within the following
region:
\bea
\lambda =[0, 1]\,, \,\,\,\,
\kappa =[-1, 1]\,,\,\,\,\,
\tan\beta = [0.2, 60]\,,\,\,\,\, 
\mu   =[-1, 1] \textrm{TeV} \,,\,\,\,\,  \nonumber \\
A_{\lambda}=[-1.0, 1.0] \textrm{TeV}\,,\,\,\,\,
A_{\kappa} =[-1.0,  1.0] \textrm{TeV}\,,\,\,\,\, \nonumber \\
M_{SUSY}   =[0.2, 3] \textrm{TeV}\,,\,\,\,\, 
M_{1}=[0.07, 3] \textrm{TeV}\,.
\eea
While varying these parameters, we take into account the experimental 
constraints on the MSSM spectrum e.g., charged Higgs mass $\ge 80$ GeV, 
chargino and scalar fermions $\ga 100$ GeV. We also apply the full set
of LEP constraints obtained from searches for neutral Higgs bosons
decaying to final states like $Z2b$, $Z4b$, $6b$, $6\tau$, $Z2b2\tau$,
$Z4\tau$, $2b2\tau$.

\begin{figure}
\begin{tabular}{cc}
\resizebox{88mm}{!}{\includegraphics{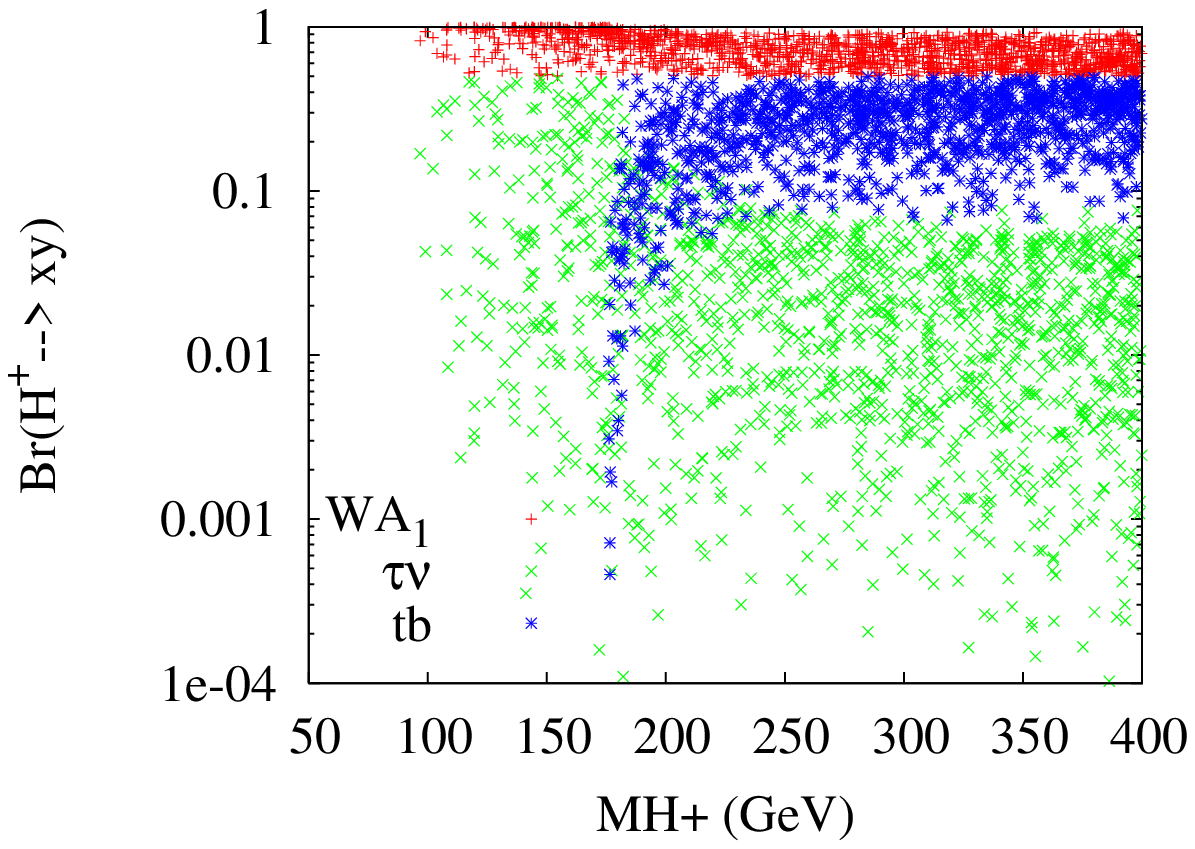}} & \hspace{-1.cm}
\resizebox{88mm}{!}{\includegraphics{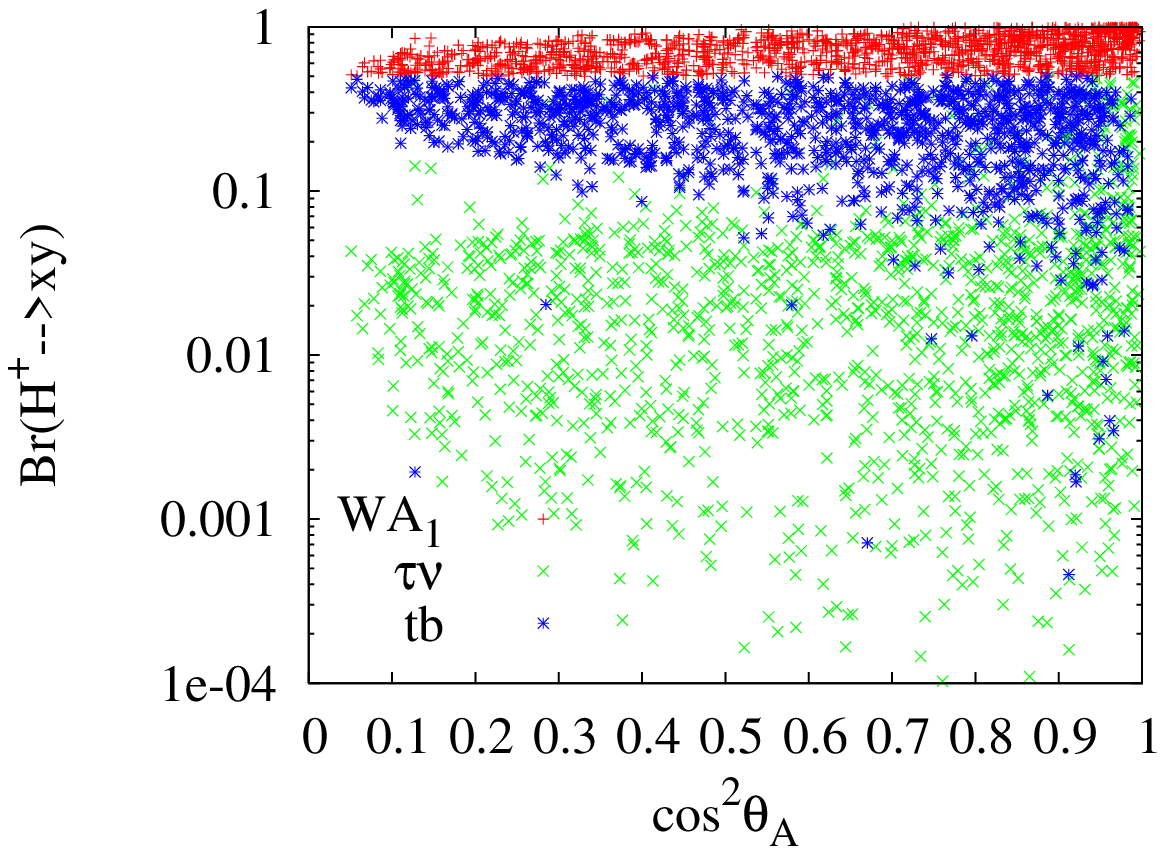}}\\
\resizebox{88mm}{!}{\includegraphics{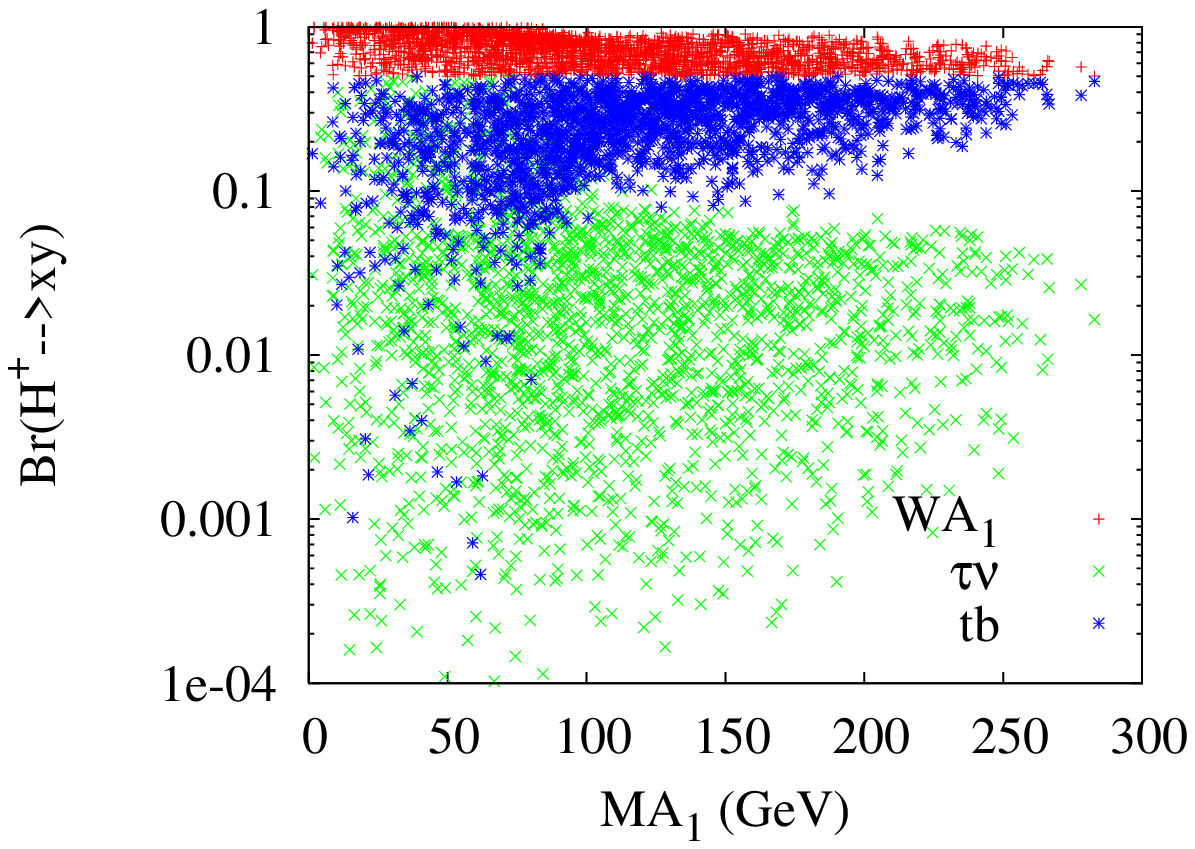}} & \hspace{-1.cm}
\resizebox{88mm}{!}{\includegraphics{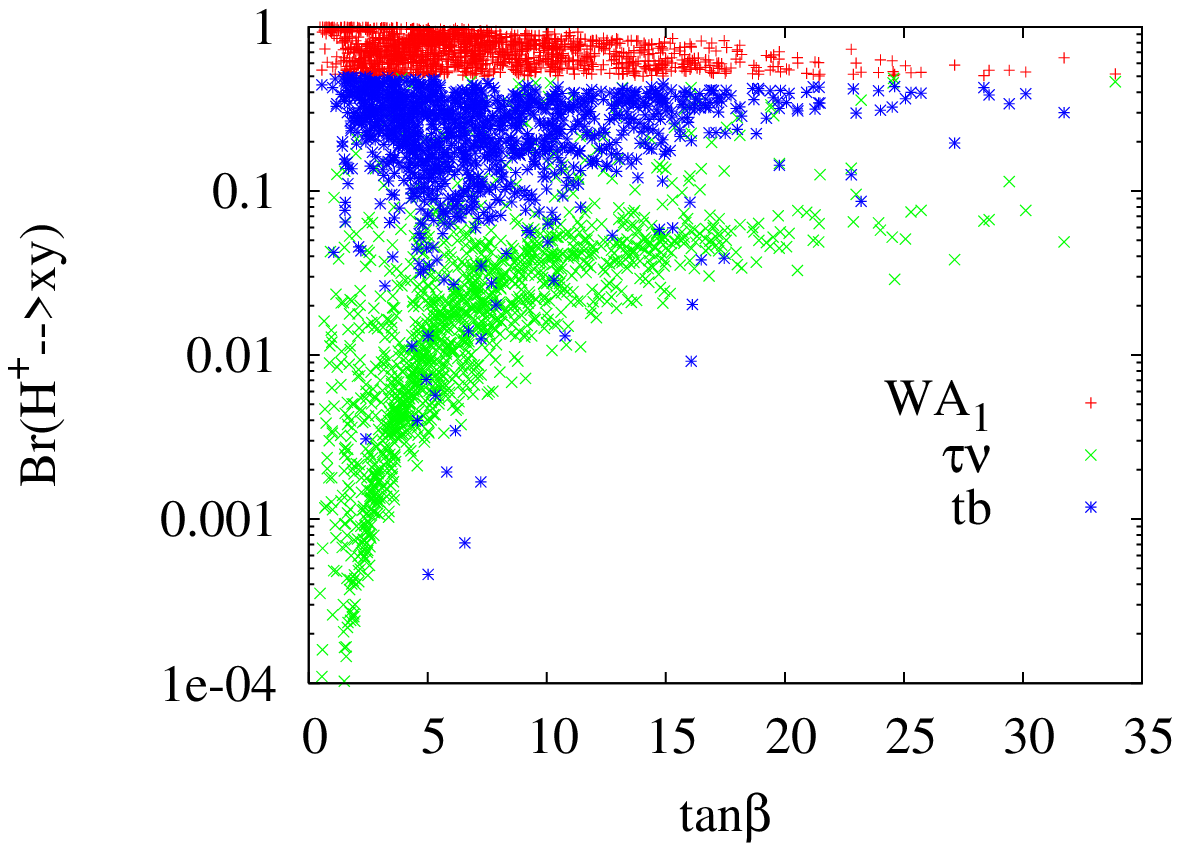}}\\
\end{tabular}
\caption
{\it Comparison of the branching ratios of $H^\pm \to \{W^\pm A_1,\tau\nu ,tb\}$
as a function of $M_{H^\pm}$ (upper left), $\cos\theta_A$ (upper right),
$M_{A_1}$ (lower left) and $\tan\beta$ (lower right).
In all panels only points with $Br(H^\pm \to  W^\pm A_1)\geq 50 \%$
are selected.}
\label{plot1}
\end{figure}

In Fig.~(\ref{plot1}) we display the branching ratios of 
$W^\pm A_1$ , $\tau\nu$  and top-bottom modes.
Before the opening of the $H^\pm \to tb$ channel,
the full dominance of $W^\pm A_1$ over $\tau\nu$ requires
light $M_{A_1} \la 100 \,GeV$, large doublet component of $A_1$ 
and $\tan\beta$ not too large. Note that at large 
$\tan\beta \approx 15-25$, the $W^\pm A_1$ and $\tau\nu$ 
channels become comparable in size.
Once the decay $H^\pm \to tb$ is open, it competes strongly with 
$W^\pm A_1$ for $\tan\beta\la 15$. As can be seen from 
Fig.~(\ref{plot1}) upper left, the branching ratio of 
$H^\pm\to W^\pm A_1$ is less than $90\%$. It is interesting
to see also that for $\cos^2\theta_A \la 0.05$ there is not a 
single point with $Br(H^\pm\to W^\pm A_1)\ga 50\% $.
Note also that at large $\tan\beta\ga 25$, it is hard for 
$H^\pm\to W^\pm A_1$ to compete with $\tau\nu$  and top-bottom modes.

\begin{figure}
\begin{tabular}{cc}
\resizebox{88mm}{!}{\includegraphics{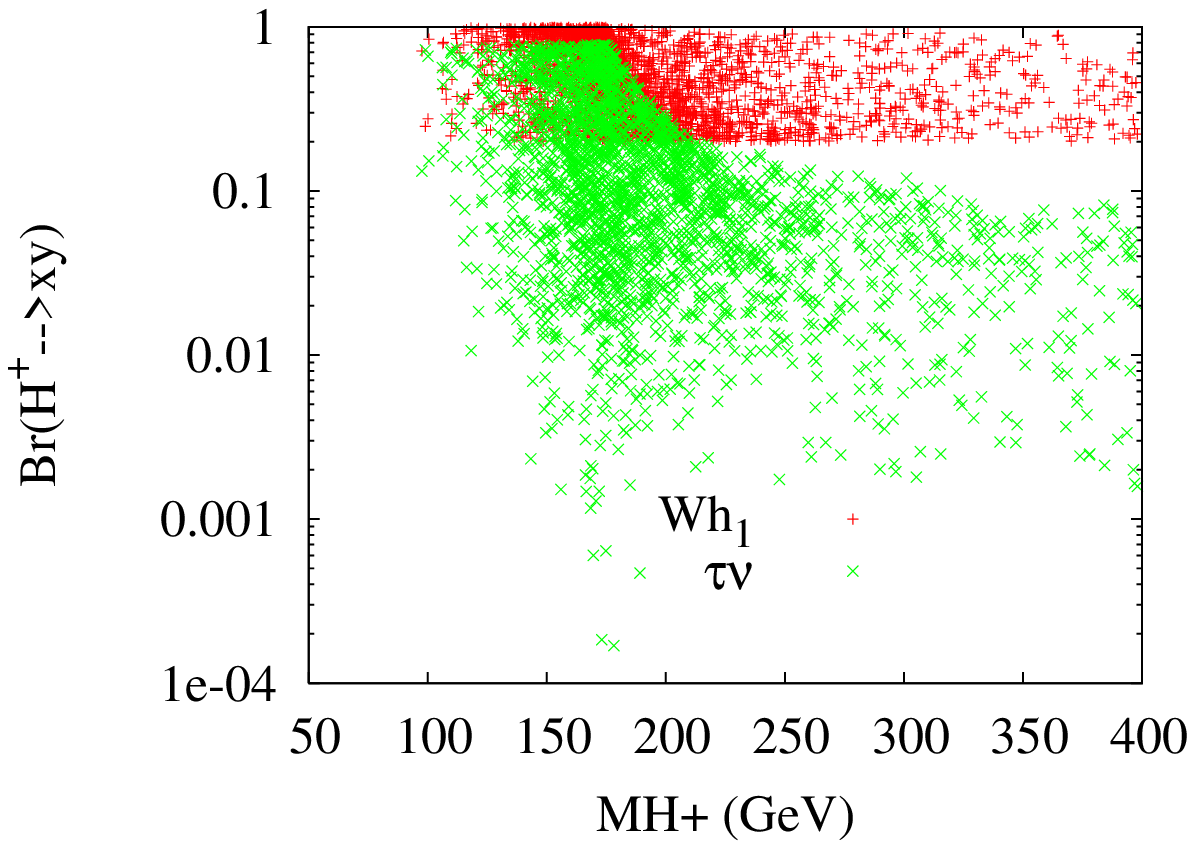}} & \hspace{-1.cm}
\resizebox{88mm}{!}{\includegraphics{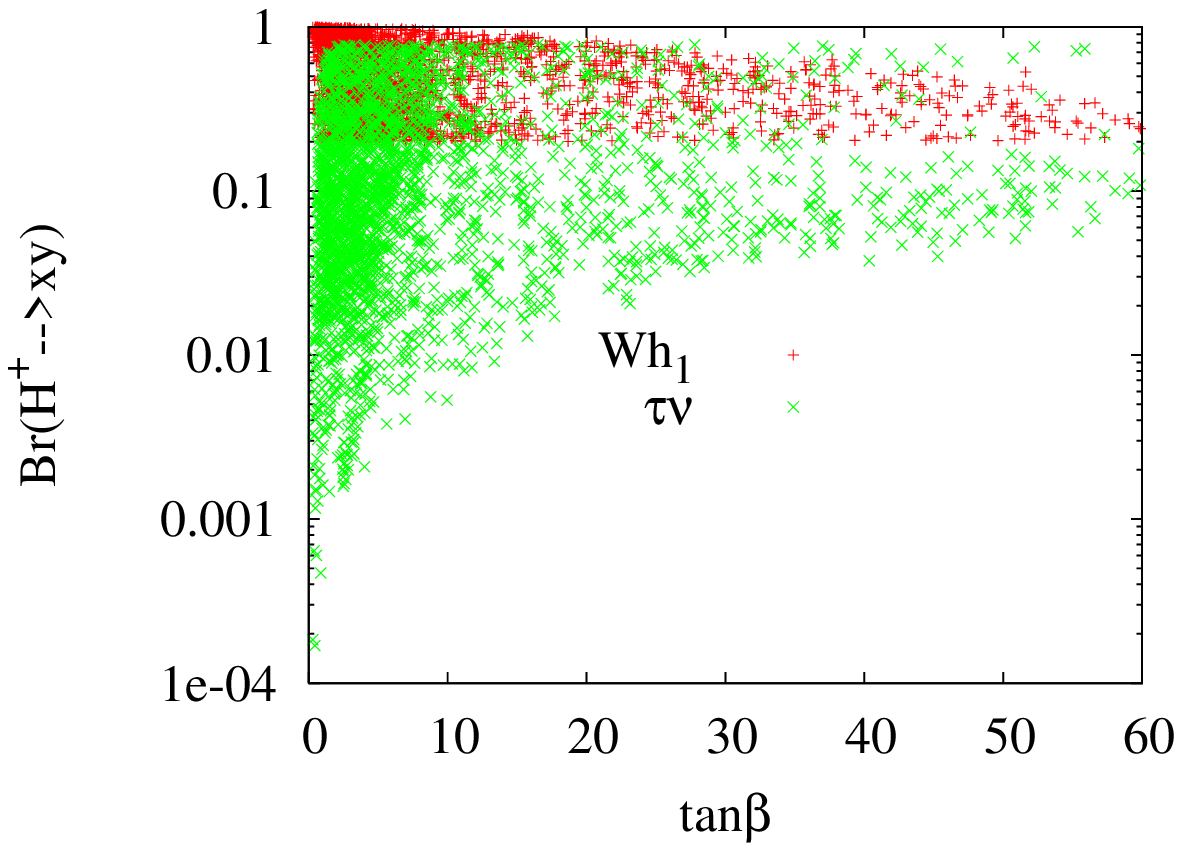}}\\
\resizebox{88mm}{!}{\includegraphics{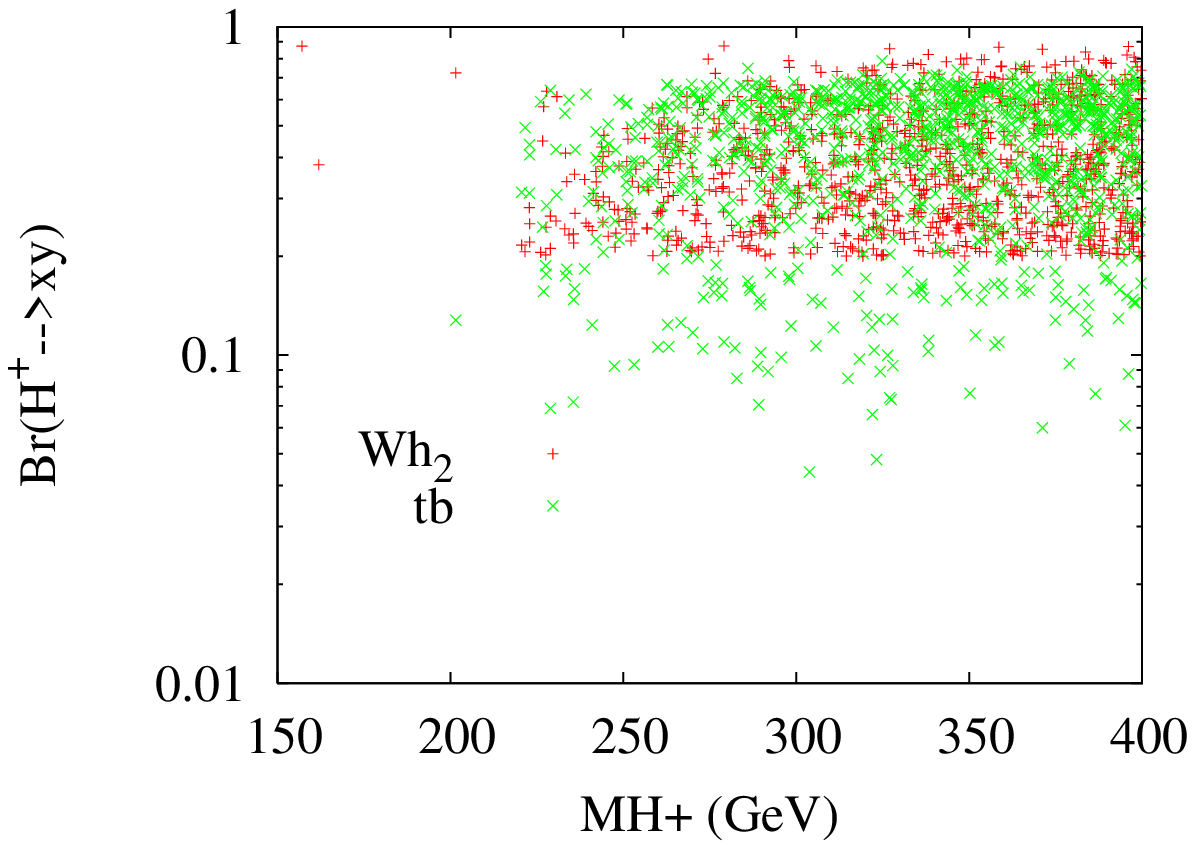}} & \hspace{-1.cm}
\resizebox{88mm}{!}{\includegraphics{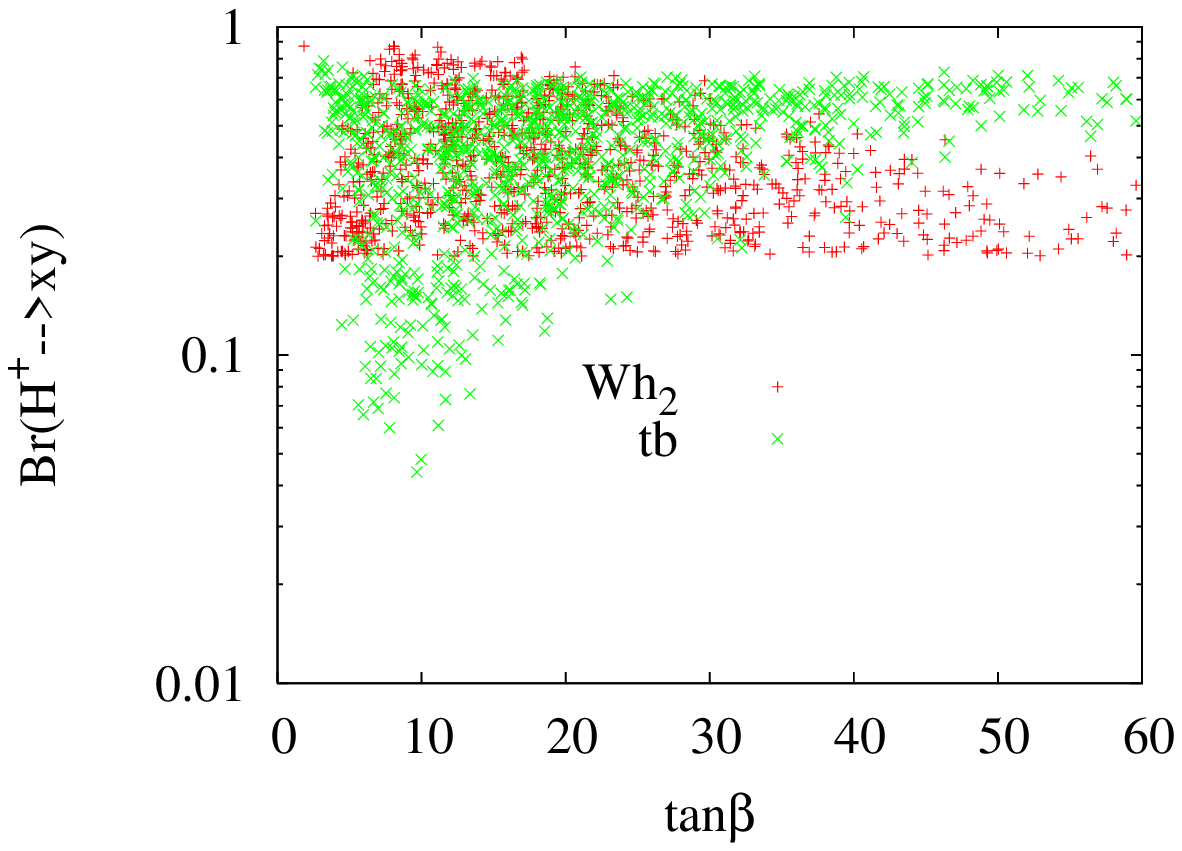}}\\
\end{tabular}
\caption
{\it Comparison of the branching ratios of $H^\pm \to \{W^\pm h_{1},\tau\nu \}$
(upper plots) and $H^\pm \to \{ W^\pm h_{2} , tb \}$ (lower plots)
as a function of $M_{H^\pm}$ (left) and $\tan\beta$ (right).
In all panels only points with $Br(H^\pm \to  W^\pm h_i)\geq 20 \%$
are selected ($i=1$ or $2$).}
\label{plot2}
\end{figure}

The case of the analogous decays $H^\pm \to W^\pm h_{1,2}$ are 
displayed in Fig.~(\ref{plot2})
as a function of $M_{H^\pm}$ and $\tan\beta$. One can see from the 
upper right 
panel of Fig.~(\ref{plot2}) that $W^\pm h_1$ 
dominates over $\tau \nu$ only for 
moderate $\tan\beta \la 5$ and before the opening of $H^\pm \to tb$ decay,
which strongly competes with $H^\pm \to W^\pm h_{1}$ mode.

>From the lower panel of Fig.~(\ref{plot2}) one can see that 
the branching ratio for $H^\pm \to W^\pm h_{2}$ can 
only be larger than
20\% for charged Higgs mass larger than about 220 GeV.
This is mainly due to the fact that $m_{h_2}$ is most of the time
larger than 140 GeV. It is clear that both $H^\pm \to W^\pm h_{2}$ 
and $H^\pm \to tb$ are of comparable size except in the case of large
$\tan\beta$ where $H^\pm \to tb$ mode dominates. \\
Importantly, we note that if $S_{13}^2\approx 1$ the 
second sum rule in Eq.~(\ref{sumrule}) requires $g_{VVh_1}^2\approx 0$ 
and $g_{W^\pm H^\mp h_1}^2\approx 0$. In this case, $S_{i3}^2\approx 1$,
both modes $H^\pm \to W^\pm h_{1,2}$  are suppressed and hence the full
dominance of $W^\pm h_{1,2}$  requires small $S_{i3}$.

In our numerical analysis we have explicitly checked that if $h_1$ is 
predominantly singlet, i.e., $S_{13}^2\ga 0.9$, 
both couplings $g_{VVh_1}^2 , g_{W^\pm H^\mp h_1}^2 \la 0.1$,
in accordance with this sum rule. 
The larger $S_{13}^2$ is, the smaller are the couplings $g_{VVh_1}^2$ and  
$g_{W^\pm H^\mp h_1}^2$. When $S_{13}^2\ga 0.9$,  
$h_1$ is almost purely singlet and even a very light $h_1$ can be allowed
by LEP experimental constraints.
In this case, both the vertices of $ZZh_1$ and $f{\bar f}h_1$ are suppressed, 
as shown in Fig. (\ref{figm2}a). In this case, the $h_2$ will be the 
Standard Model like Higgs boson and the coupling $g_{VVh_2}$ can be large, 
as indicated 
by the first sum rule in Eq. (\ref{sumrule}) and demonstrated 
in Fig. (\ref{figm2}a). 

In the converse case when $S_{13}^2 \approx 0\to 0.1$,
$g_{VVh_1}^2$ and $g_{W^\pm H^\mp h_1}^2$ have to 
share the quantity $1-S_{13}^2$. Since $h_1$ is 
dominantly doublet the coupling $g_{W^\pm H^\mp h_1}$ can be maximal,
and hence the branching ratio  of $H^\pm \to W^\pm h_1$ can be large.

\subsection{ The cross-sections for $pp \rightarrow H^\pm h_1$, 
$pp \rightarrow W^\pm h_1$  and $pp \rightarrow H^\pm A_1$
in the NMSSM }

Searches for Higgs bosons at the LHC suffer from 
large QCD backgrounds. However, detailed studies have shown that
multiple signals for the MSSM Higgs bosons 
are possible in a sizeable region of the plane [$\tan\beta, M_{H^\pm}]$
\cite{Assamagan:2002ne}.
Much of these studies for the MSSM can be applied to the NMSSM 
with some caveats which were discussed in Section IV.
The most problematic region for $H^\pm$ discovery in the MSSM
is for moderate values 
of $\tan\beta$, since the production mechanisms which rely on a large 
bottom quark or top quark Yukawa coupling (e.g. $gb \to H^\pm t$)
are least effective.
Hence alternative mechanisms which could offer good detection 
prospects for $H^\pm$ at moderate values of $\tan\beta$ are desirable.

\begin{figure}
\begin{tabular}{cc}
\resizebox{88mm}{!}{\includegraphics{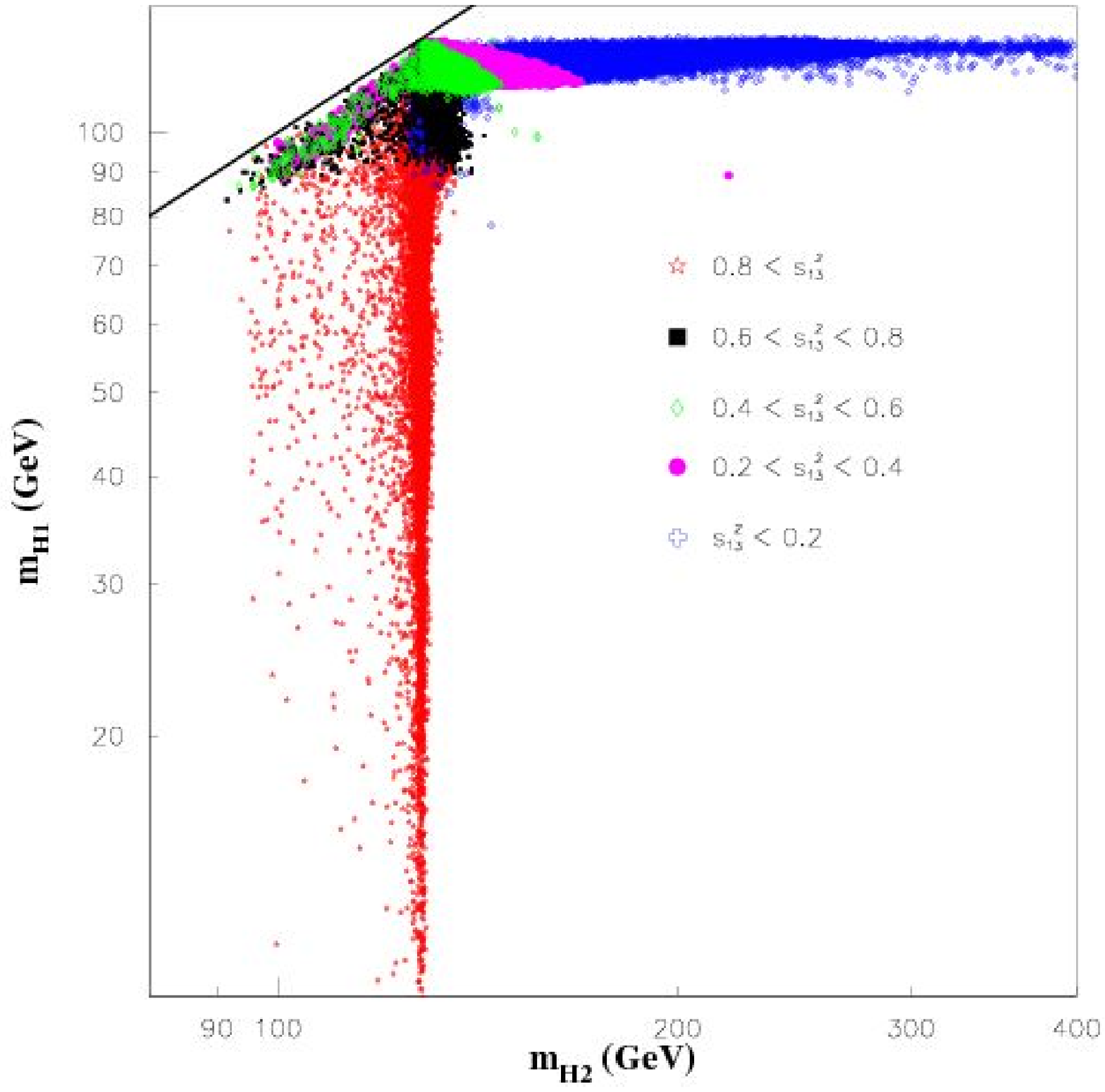}} & \hspace{-1.cm}
\resizebox{88mm}{!}{\includegraphics{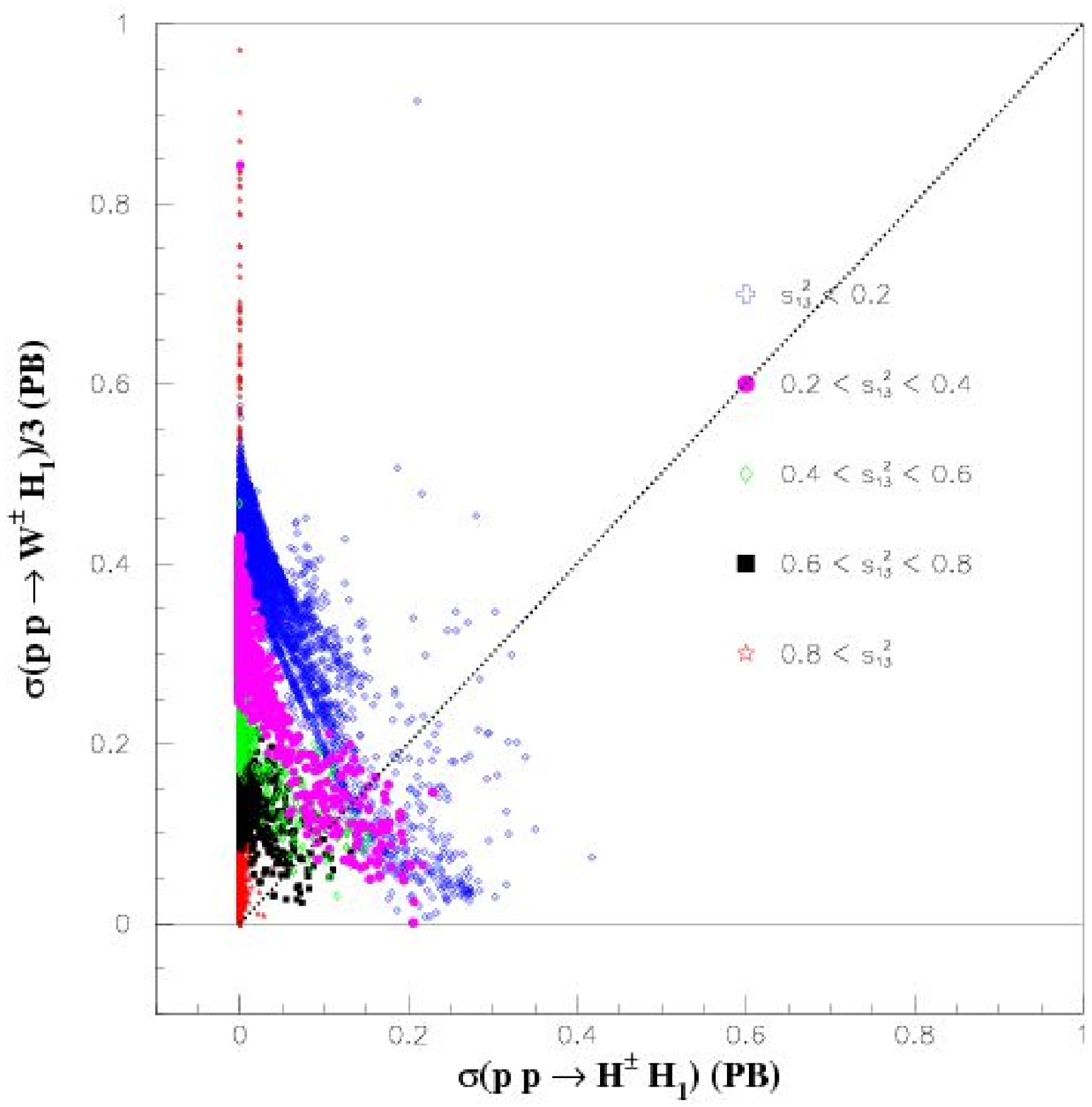}}
\end{tabular}
\caption{\it  Left panel: the comparison of $m_{h_1}$ and $m_{h_2}$ with 
respect to $S_{13}^2$. Right panel: the comparison of the cross section of 
the processes $pp\to W^\pm h_1$ and $pp \to H^\pm h_1$.For 
$\sigma(pp \to H^\pm H_1)$, we sum over $\sigma(pp \to H^+ H_1)$ and 
$\sigma(pp \to H^- H_1)$, while for $\sigma(pp \rightarrow W^\pm h_1)$ 
we sum over $\sigma(pp \to W^+ h_1)$ and $\sigma(pp \to W^- h_1)$. 
For the sake of comparison, we deliberately divide $\sigma(pp \to W^\pm h_1)$ 
by three due to the three helicity states of massive vector boson $W$. }
\label{figm2}
\end{figure}

The cross sections for the
pair production mechanisms $pp \rightarrow H^\pm A_1$ and
$pp \rightarrow H^\pm h_1$ fall quickly with increasing
scalar masses but for relatively light masses 
($\la 200$ GeV) they can provide promising signal rates
which might enable their detection at the LHC. 
One common feature is that the produced scalars
enjoy large transverse momenta, which are crucial for the 
trigger and event selection. 
The cross section for $pp\to W^\pm \to H^\pm A$ was first studied
\cite{Kanemura:2001hz} at both the LHC and
Tevatron in the CP conserving MSSM for $M_A> 100$ GeV. 
The analogous process $pp\to H^\pm h_1$ for a very light $h_1$ with
unsuppressed coupling $h_1 H^\pm W^\mp$ was studied
in the 2HDM and the CP violating MSSM in \cite{Akeroyd:2003jp}
at the Tevatron. In \cite{Belyaev:2006rf} it was shown that 
$pp\to H^\pm h^0,H^\pm A^0$ can be important in specific regions of parameter 
space (i.e., very light $h^0$, $A^0$) in the CP conserving MSSM. 

In the NMSSM, if the coupling $H^\pm W^\mp A_1$ is  sizeable,
so will be the cross section for $pp\to W^\pm \to H^\pm A_1$
provided that $H^\pm$ and  $A_1$ are not too heavy.
The production mechanism $pp\to H^\pm A_1$ 
followed by the decay $H^\pm\to W^\pm A_1$  
would give rise to a signal $W^\pm A_1A_1\to Wbbbb$ 
\cite{Akeroyd:2003jp} or $W^\pm A_1A_1\to W\tau\tau\tau\tau$.
The signature $W^\pm A_1A_1\to Wbbbb$ was simulated at the 
LHC in \cite{Ghosh:2004wr}
in the context of the CP violating MSSM with the conclusion that a 
sizeable signal essentially free of background could be obtained.
We use NMSSM-TOOLS1.1.1 to calculate the mass spectrum and 
couplings of the NMSSM
Higgs bosons, and we link CTQ6.1M PDF distribution to this code 
in order to calculate the cross sections of $pp \rightarrow H^\pm A_1$,
$pp \rightarrow H^\pm h_1$ and $pp \rightarrow W^\pm h_1$.
All cross sections are evaluated at a scale which is the sum of the masses
in final states and do not include
next-to-leading order QCD enhancement factors (K factors) of around $1.2
\to 1.3$ \cite{Kanemura:2001hz},\cite{Han:1991ia}.

For our numerical analysis, we have done a systematic
scan with NMSSM-TOOLS1.1.1 \cite{Ellwanger:2004xm}. Firstly, we explore 
the phenomenological implication of the sum rule Eq. (\ref{sumrule}) 
with Fig. (\ref{figm2}b). 
There are several comments in order:\\
1) All points in Fig. (\ref{figm2}b) respect the following constraint 
$M_{H^\pm}\ga M_W$, this leads to a 
smaller cross section for $\sigma(pp \to H^\pm h_1)$.\\
2) As $S_{13}^2$ increases both processes are
suppressed due to the decrease of the couplings $W^\mp W^\pm h_1$ 
and $W^\mp H^\pm h_1$.\\
3) When $S_{13}^2>0.8$, $h_1$ is dominated by singlet component, therefore it 
can be very light see Fig. (\ref{figm2}a). In this cases, according to 
sum rule Eq. (\ref{sumrule}), the vertex $VVh_1$ suffers a severe suppression. 
However, some points with large $\sigma(pp \to W^\pm h_1)$ arise
due to the fact that a very light $h_1$ is allowed.

In Fig. ({\ref{figm1}a) we study the cross section of 
$pp \rightarrow H^\pm h_1$ at the LHC and select points which 
simultaneously satisfy the following conditions:
\begin{eqnarray}
\sigma(pp \rightarrow H^\pm h_1) > 0.1 \ pb \, \qquad \rm{and} \qquad
Br(H^\pm \to W^\pm A_1) > 0.5 \,.
\label{cond-ch}
\end{eqnarray}
We require points in parameter space with cross sections larger 
than $0.1$ pb as a conservative threshold of observability 
for this channel at the LHC. From the figure it is clear
that $\sigma(pp \rightarrow H^\pm h_1)< 0.5$ pb
when the charged Higgs boson decays dominantly to $W^\pm A_1$.

In Fig. (\ref{figm1}b), we study the cross section of 
$pp \rightarrow W^\pm h_1$ at LHC and select points which 
satisfy the following conditions:
\begin{eqnarray}
\sigma(pp \rightarrow W^\pm h_1) > 0.1\,pb \, \qquad \rm{and} \qquad
Br(h_1 \to A_1  A_1)  >  0.5 \,.
\label{cond-vvh}
\end{eqnarray}
The typical cross section for $\sigma(pp \rightarrow W^\pm h_1)$ 
is around a few pb, which is considerably larger than
$\sigma(pp \rightarrow H^\pm h_1$). The larger cross sections
correspond to the larger branching ratios for 
$h_1\to A_1 A_1$ $(> 90\%$).
The numerical results in  Fig. (\ref{figm1}b) are in good agreement 
with analogous results presented in \cite{Moretti:2006hq}.

\begin{figure}
\begin{tabular}{cc}
\resizebox{88mm}{!}{\includegraphics{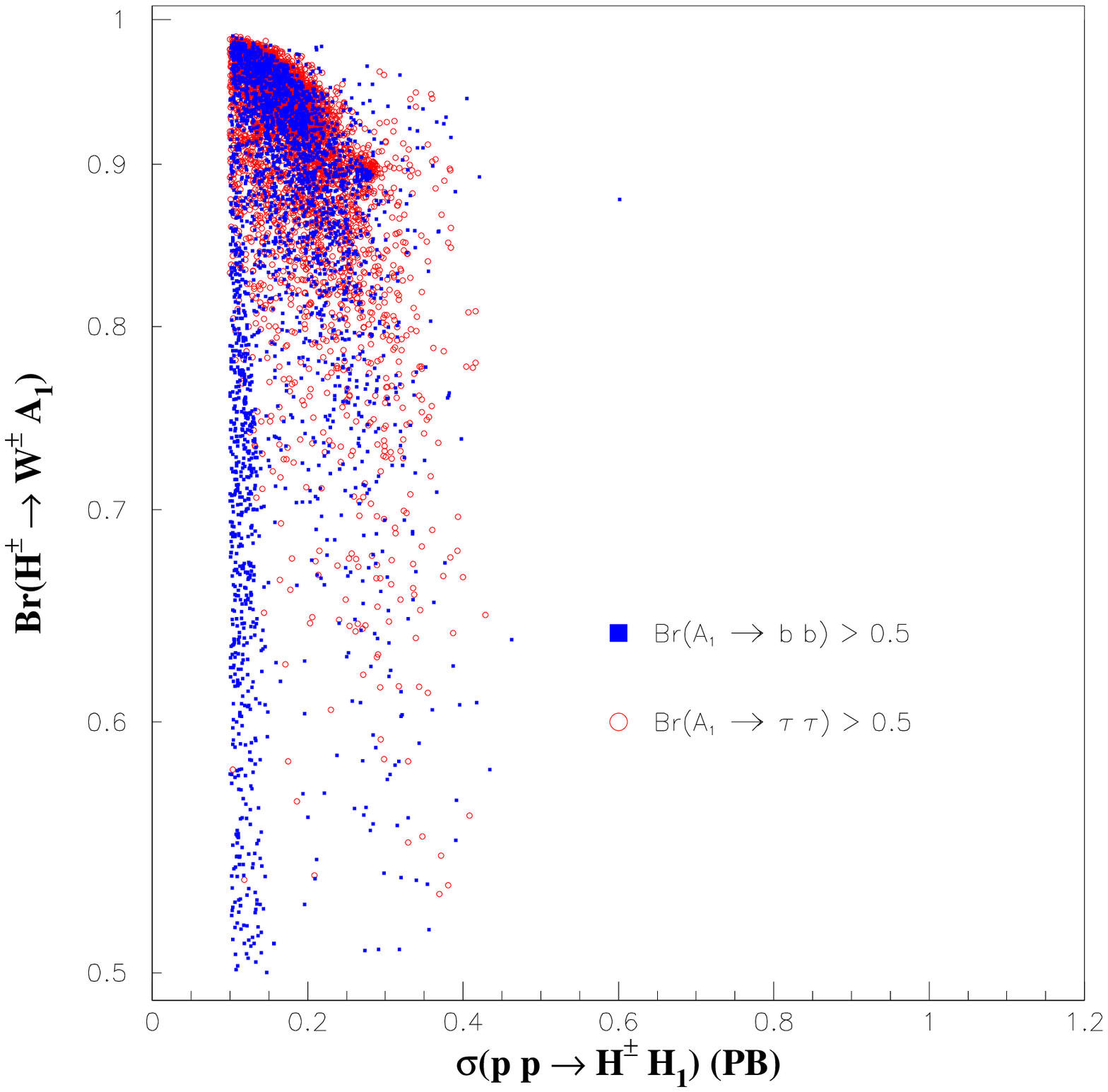}} & \hspace{-1.cm}
\resizebox{88mm}{!}{\includegraphics{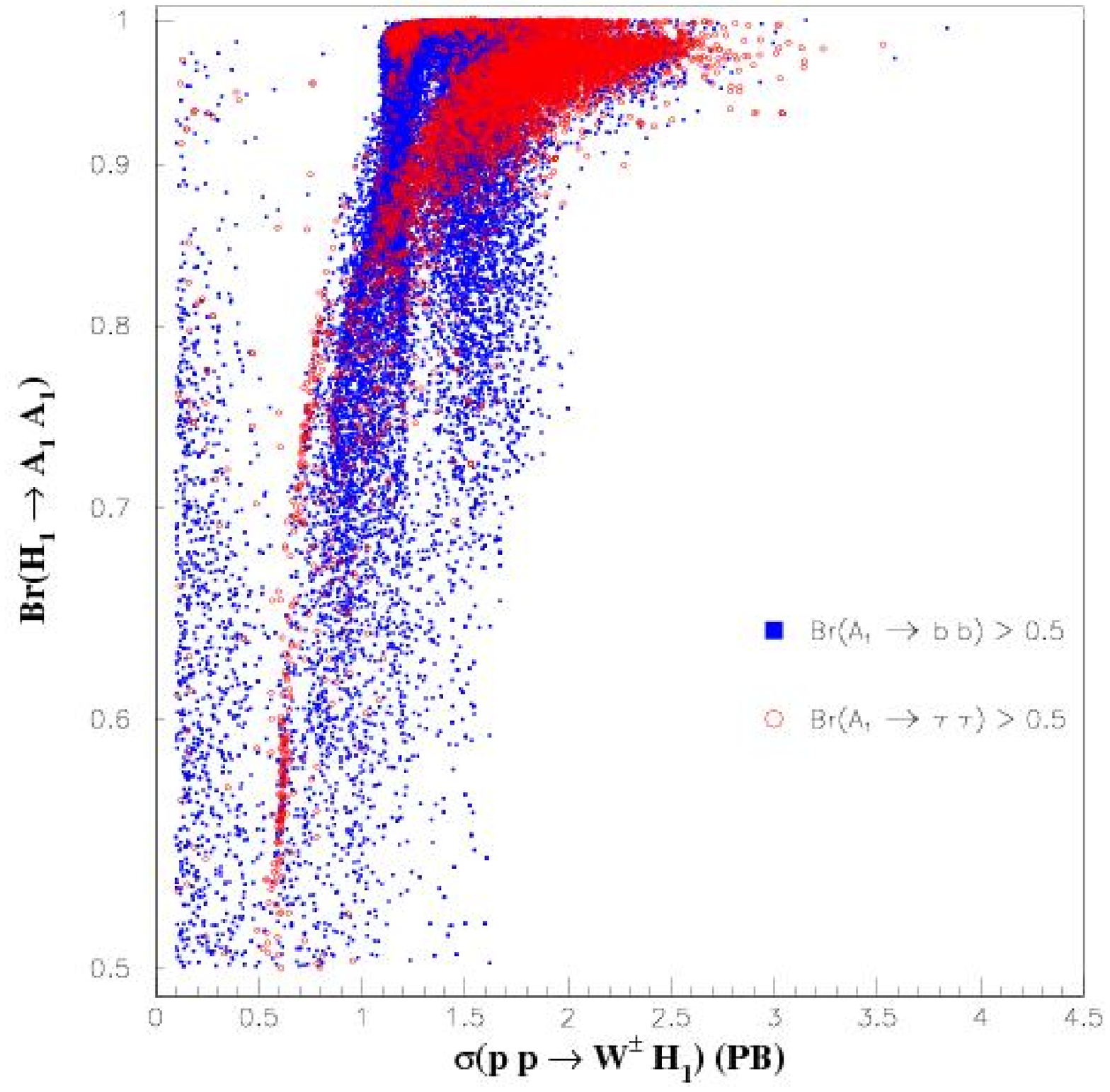}}
\end{tabular}
\caption{\it Left panel: points selected with the condition given in 
Eq. (\ref{cond-ch}). Right panel: points selected with the condition given 
in Eq. (\ref{cond-vvh}). For $\sigma(pp \rightarrow H^\pm h_1)$ we sum over
$\sigma(pp \rightarrow H^+ h_1)$ and  $\sigma(pp \rightarrow H^- h_1)$;
for $\sigma(pp \to W^\pm h_1)$ we sum over $\sigma(pp \rightarrow W^+ h_1)$ and
$\sigma(pp \to W^- h_1)$. We show the two decay modes of 
$A_1$: $A_1 \to b {\bar b}$,  and $A_1 \to \tau {\bar \tau}$,  
which corresponds to two mass regions: $2 m_b <M_{A_1}<m_{h_1}/2$, and 
$2 m_\tau <M_{A_1}< 2 m_b$, respectively. }
\label{figm1}
\end{figure}

\begin{figure}
\begin{tabular}{cc}
\resizebox{88mm}{!}{\includegraphics{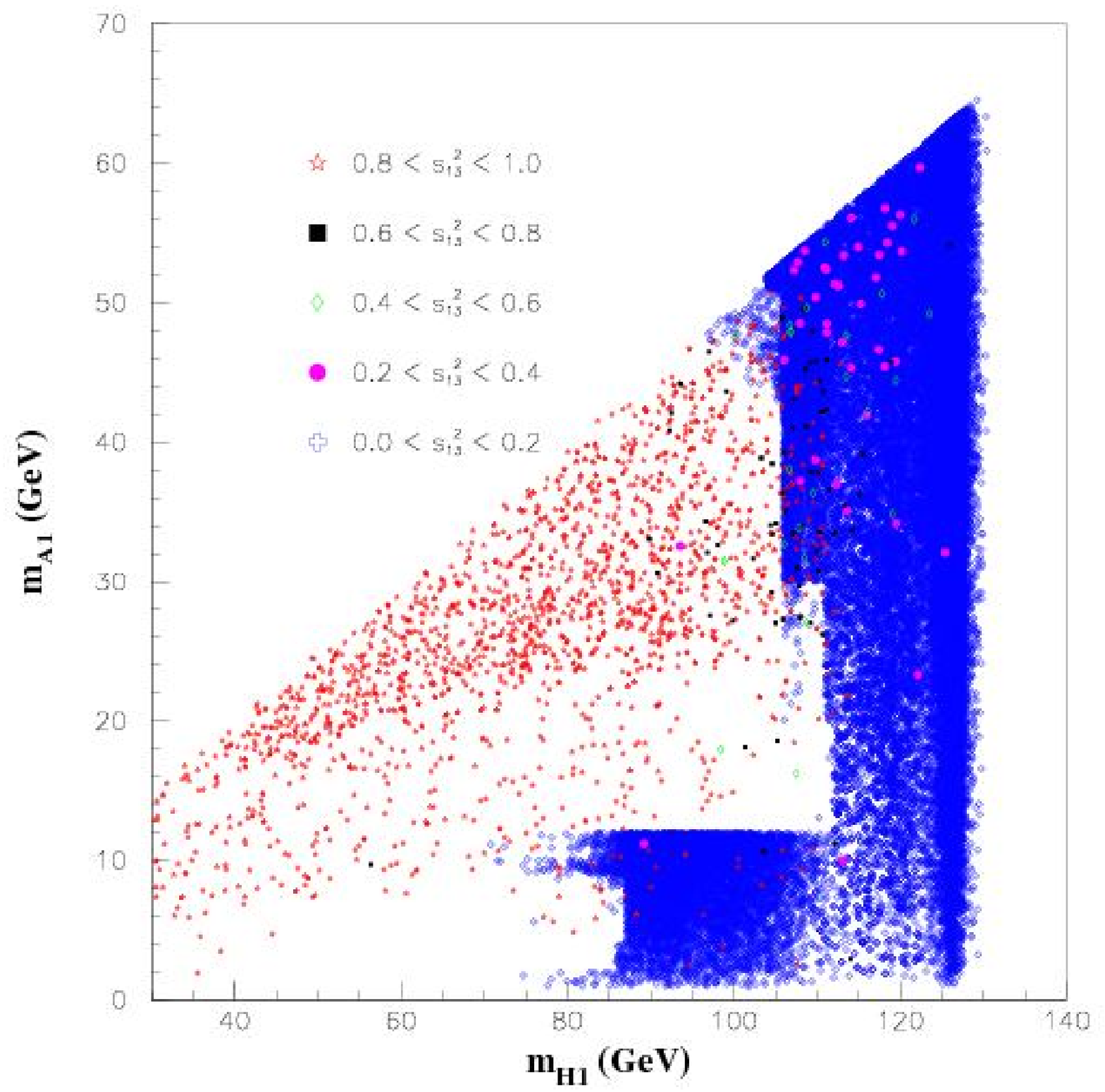}} & \hspace{-1.cm}
\resizebox{88mm}{!}{\includegraphics{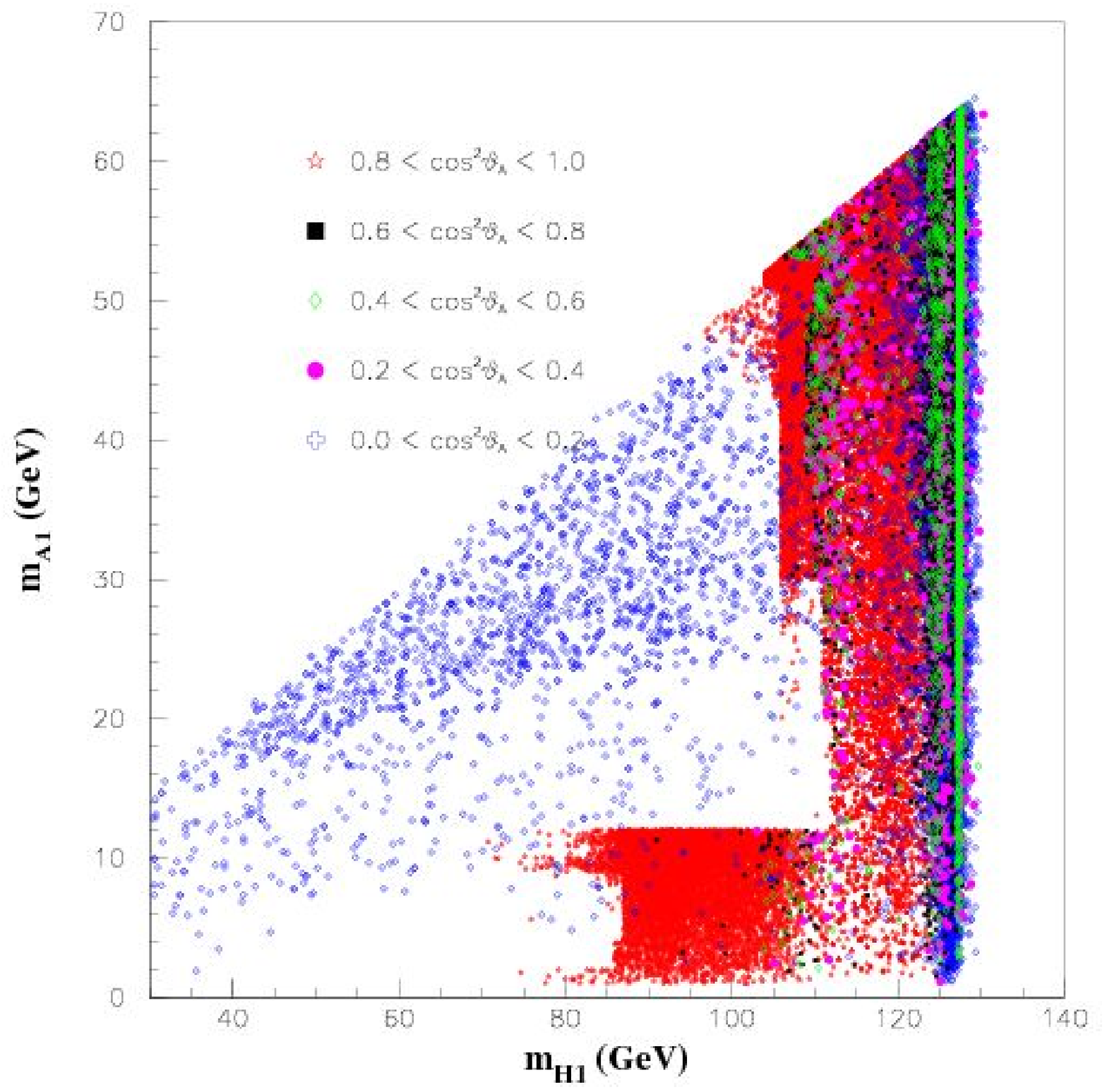}}
\end{tabular}
\caption{\it Parameter space satisfying $Br(H_1 \to A_1 A_1) \ge 0.5$
in the plane $[m_{h_1},m_{A_1}]$.  
The components of both $h_1$ and $A_1$ are displayed.}
\label{fig0}
\end{figure}

In Fig. (\ref{fig0}) we analyze the components of $H_1$ and $A_1$.
Points which satisfy the following condition are selected:
\bea
Br(h_1 \to A_1 A_1) \ge 0.5\,.
\eea
As expected, when both $h_1$ and $A_1$ are dominantly composed of 
doublet fields the region of light Higgs bosons is ruled out
from searches for $e^+ e^- \to Z h_1 \to Z 2 A_1 \to Z 4 b$,
and $m_{h_1}$ should be heavier than around $100 \sim 110$ GeV. 
When $M_{A_1} \la 2 m_b$, $m_{h_1}$ can be lighter than $100$ GeV
due to the fact that the LEP2 sensitivity to the 
channel $e^+ e^- \to Z 4 \tau$ was less robust 
than that for $e^+ e^- \to Z 4b$.
Interestingly, when both $h_1$ and $A_1$ are mainly singlet and hence 
the vertex of $VV h_1$ is greatly suppressed, much lighter values for 
$m_{h_1}$ ($\la 80$ GeV) are still allowed, as shown by points with red 
stars in Fig. (\ref{fig0}a) and blue crosses in Fig. (\ref{fig0}b). 

This process $pp\to H^\pm A_1\to W^\pm A_1A_1$ leads to the same
signature as the process $pp \to Wh_1 \to WA_1A_1\to Wbbbb$.
The latter has been simulated in \cite{Cheung:2007sv} and also offers
very good detection prospects. We will compare the magnitude of these two
distinct mechanisms which lead to the same $Wbbbb$ signature. In addition, 
the mechanism $pp\to H^\pm h_{1}$ followed by the decay $H^\pm\to W^\pm A_1$ 
would also lead to the same final state $W^\pm A_1h_1\to Wbbbb$.
We will concentrate on the scenario where $h_1\to A_1A_1$ is large 
and thus $h_1 \to b\bar{b}$ will be kinematically suppressed. We will 
discuss the magnitude of $pp\to H^\pm h_{1}\to W^\pm A_1h_{1}\to Wbbbb$ later.

In Fig. (\ref{fig1}a) we study the process 
$pp \rightarrow H^\pm A_1$ by choosing points
which satisfy the following conditions:
\begin{eqnarray}
\sigma(pp \rightarrow H^\pm A_1) > 0.1\ pb \, \qquad \rm{and} \qquad
Br(H^\pm \to W^\pm  A_1) > 0.5 \,.
\label{cond-ca}
\end{eqnarray}
It is apparent that the magnitude of $\sigma(pp \rightarrow H^\pm A_1$)
can reach a few pb and thus is within the detection capability of 
the LHC.
The analysis of \cite{Ghosh:2004wr} (for the CP violating MSSM) 
suggests that $\sigma(pp \rightarrow H^\pm A_1) \ga 0.1$ pb with 
a large $Br(H^\pm \rightarrow W^\pm A_1)$ would 
be sufficient for an observable $Wbbbb$ signal at the LHC.
Most strikingly, the cross section of the process
$pp \rightarrow H^\pm A_1$ can be comparable to 
that of $pp \to W^\pm h_1$.

\begin{figure}
\begin{tabular}{cc}
\resizebox{88mm}{!}{\includegraphics{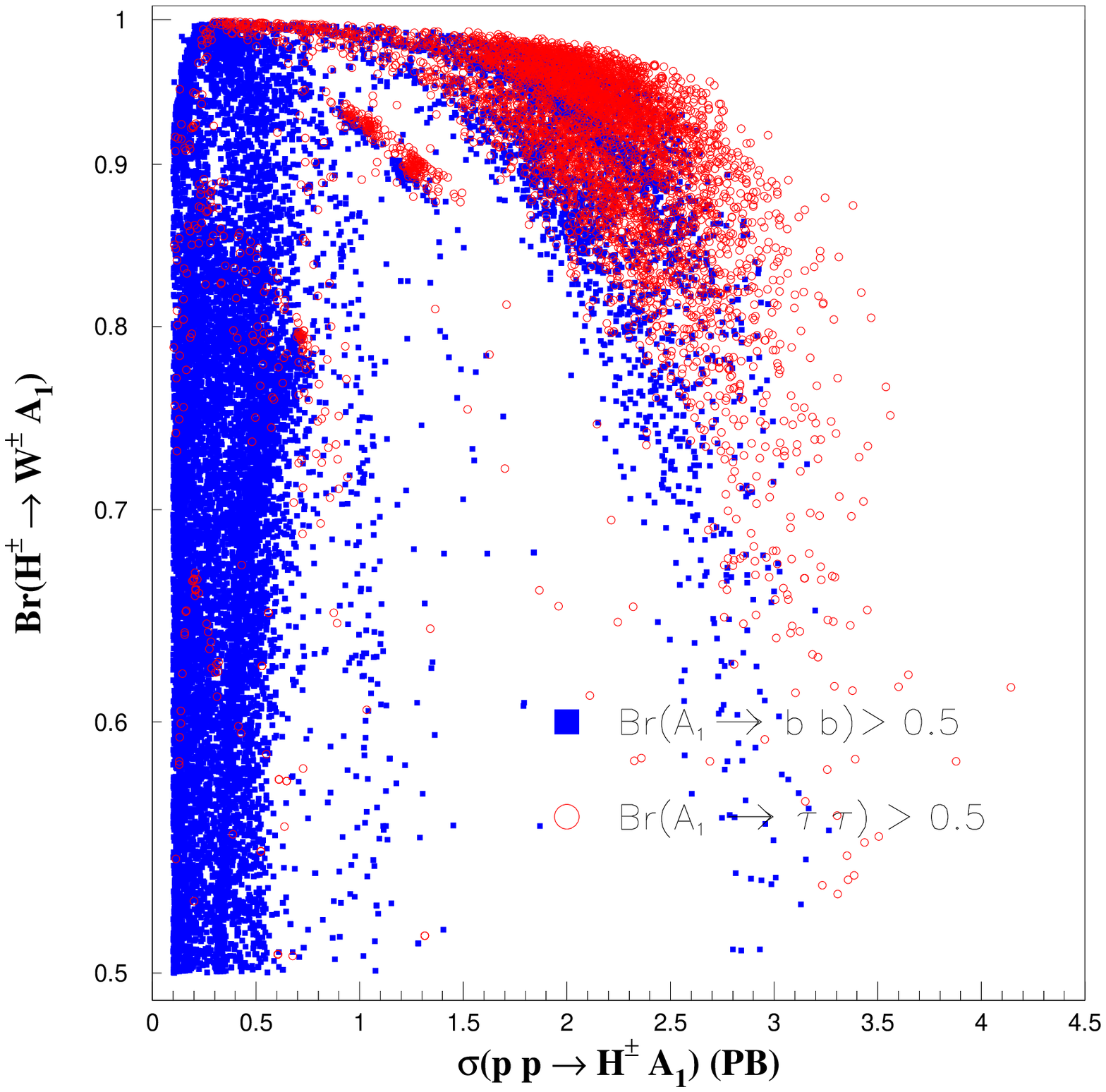}} & \hspace{-1.cm}
\resizebox{88mm}{!}{\includegraphics{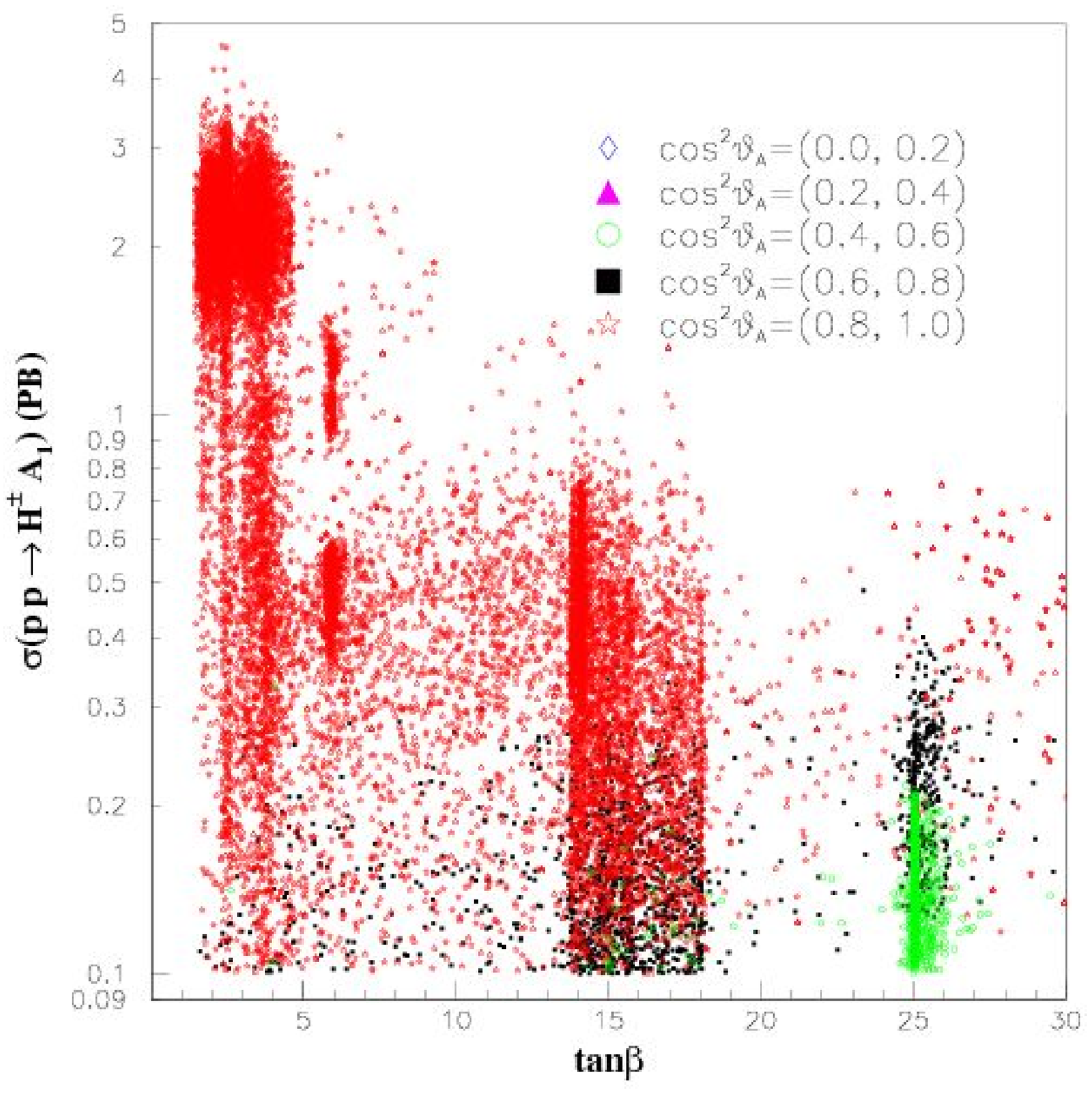}}
\end{tabular}
\caption{\it  
Left panel: Points in  the plane 
$[\sigma(pp \rightarrow H^\pm A_1),Br(H^\pm\to A_1W]$ which
satisfy the conditions given in Eq. (\ref{cond-ca}).
For $\sigma(pp \rightarrow H^\pm A_1)$ we sum over
$\sigma(pp \rightarrow H^+ A_1)$ and $\sigma(pp \rightarrow H^- A_1)$. 
We show the two decay modes of $A_1$: $A_1 \to b {\bar b}$  and 
$A_1 \to \tau {\bar \tau}$, which corresponds to two mass regions: 
$2 m_b <M_{A_1}<m_{h_1}/2$, and $2 m_\tau <M_{A_1}< 2 m_b$, respectively.
Right panel: the dependence of $\sigma(pp \rightarrow H^\pm A_1)$
on both $tan\beta$ and $cos\theta_A$ are displayed.
}
\label{fig1}
\end{figure}

The majority of the points in Fig. (\ref{fig1}a) correspond to the 
parameter space where $\tan\beta$ is located in the 
range $0.2 \la \tan\beta \la 20$. 
As seen in the previous section,  when $\tan\beta \ga 20$ 
the decay channel $H^\pm \rightarrow \tau^\pm \nu_{\tau}$ (or $H^\pm \to tb$ )
will dominate over $H^\pm \rightarrow W^\pm A_1$.
It is evident from  Fig. (\ref{fig1}a) that there are
plenty of points with $\sigma(pp \rightarrow H^\pm A_1) \ga 0.1$ pb  
and $Br(H^\pm \rightarrow W^\pm A_1)\ga 90 \%$.

\begin{figure}
\begin{tabular}{cc}
\resizebox{88mm}{!}{\includegraphics{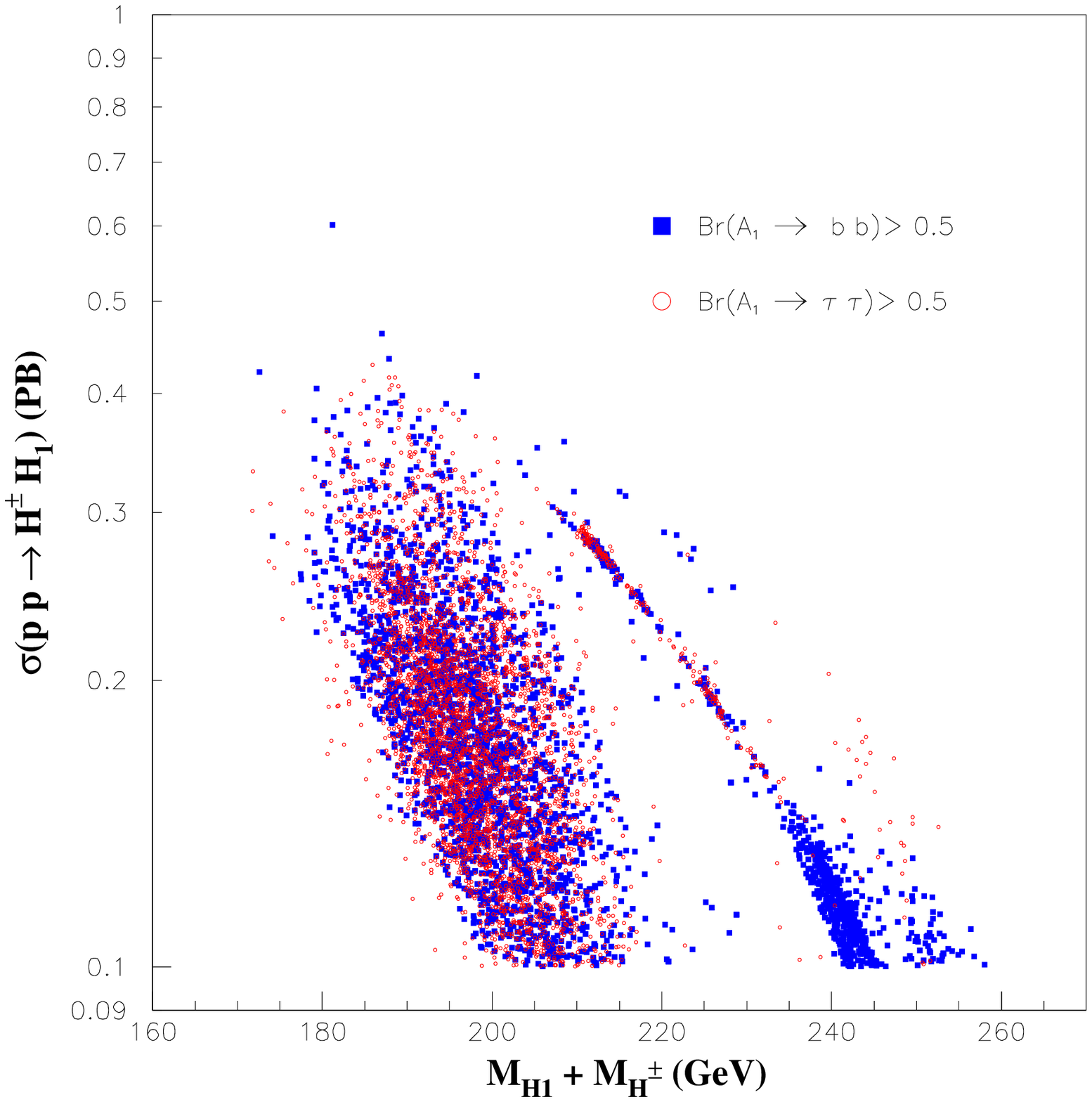}} & \hspace{-1.cm}
\resizebox{88mm}{!}{\includegraphics{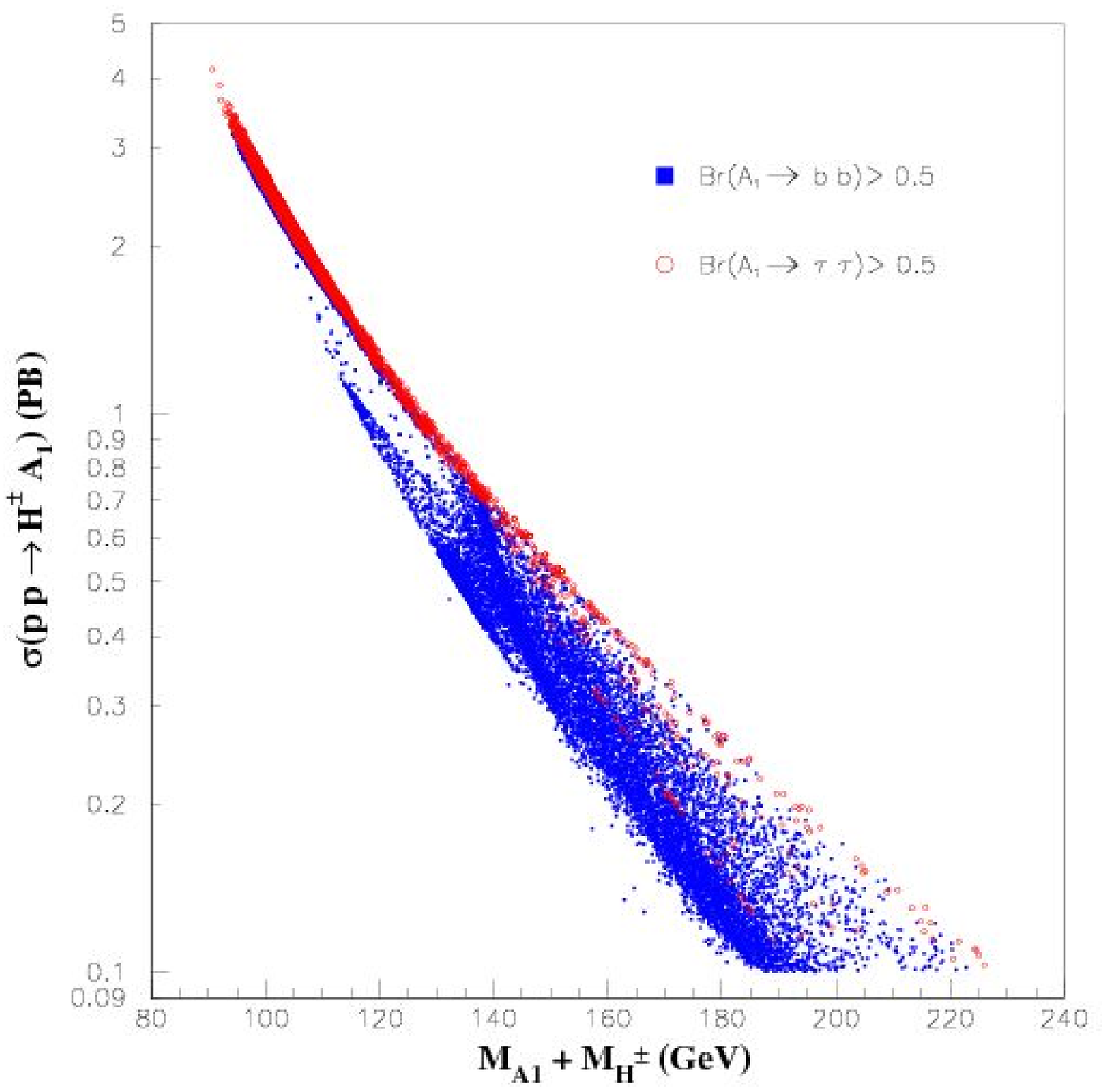}}
\end{tabular}
\caption{\it Left panel: the cross section of $pp \to H^\pm h_1$ against 
$M_{h_1}+M_{H^\pm}$; all points satisfy the condition in Eq. (\ref{cond-ch}). 
Right panel: the cross section of $pp \to H^\pm A_1$ against 
$M_{A_1} + M_{H^\pm}$; all points satisfy the condition in Eq. (\ref{cond-ca}). 
We show the two decay modes of $A_1$: $A_1 \to b {\bar b}$ and 
$A_1 \to \tau {\bar \tau}$,  which corresponds to two mass regions: 
$2 m_b <M_{A_1}<m_{h_1}/2$, and $2 m_\tau <M_{A_1}< 2 m_b$, respectively.}
\label{fig2}
\end{figure}

In Fig. (\ref{fig1}b) we show the dependence of 
$\sigma(pp \rightarrow H^\pm A_1)$ on  $tan \beta$ and $cos \theta_A$. 
The figure clearly shows that when $A_1$ is mainly doublet the cross section 
$\sigma(pp \rightarrow H^\pm A_1)$ can reach a few $pb$. 
Importantly, the cross section can be sizeable in the whole
region $1 < tan\beta < 30$, and thus this mechanism can be applied
to the region $ 5 \la tan\beta \la 20$ for which
$H^\pm$ discovery in the conventional production mechanisms (which utilize 
the $t$ and $b$ quark Yukawa couplings) are least effective.
Thus $H^\pm$ production 
via $pp \rightarrow H^\pm A_1$ might offer the best prospects for 
the detection of a light NMSSM charged Higgs boson 
in the region of intermediate  $\tan\beta$. 
It is clear from Fig. (\ref{fig1}b) that there are no 
points at all with $0\la cos^2 \theta_A \la 0.4$, the reason being that
such points do not satisfy the requirement 
$\sigma(pp \rightarrow H^\pm A_1) \ga 0.1$ pb.

Fig. (\ref{fig2}) shows the dependence of 
$\sigma(pp \rightarrow H^\pm h_1 (A_1))$ on $m_{h_1} (m_{A_1}) + m_{H^\pm}$. 
Points in Fig. (\ref{fig2}a) satisfy the conditions given in 
 Eq. (\ref{cond-ch}), while points in Fig. (\ref{fig2}b) satisfy the 
conditions given in Eq. (\ref{cond-ca}). Clearly the points with large
cross section correspond to the region in the parameter space where
both $H^\pm$ and $h_1(A_1)$ are light and the couplings $W^\mp H^\pm h_1$
and $W^\mp H^\pm A_1$ are near maximal. In Fig. (\ref{fig2}b),
it is evident that the cross section 
for $pp \rightarrow H^\pm A_1$ can reach a few pb
when $A_1$ is as light as 10 GeV. In contrast, in
Fig. (\ref{fig2}a) one can see that that there are only points for 
$m_{h_1}+ m_{H^\pm}\ga 170$ GeV which corresponds to $m_{H^\pm}\ga 80$
GeV and $m_{h_1}\ga 90$ GeV. The lack of sample points with 
large cross section for $pp \rightarrow H^\pm h_1$ 
is due to difficulties in finding points with relatively light
$h_1$ and $H^\pm$ (i.e., $m_{h_1}+ m_{H^\pm}\la 170$ GeV)
which can satisfy the experimental constraints.

\begin{figure}
\begin{tabular}{cc}
\resizebox{88mm}{!}{\includegraphics{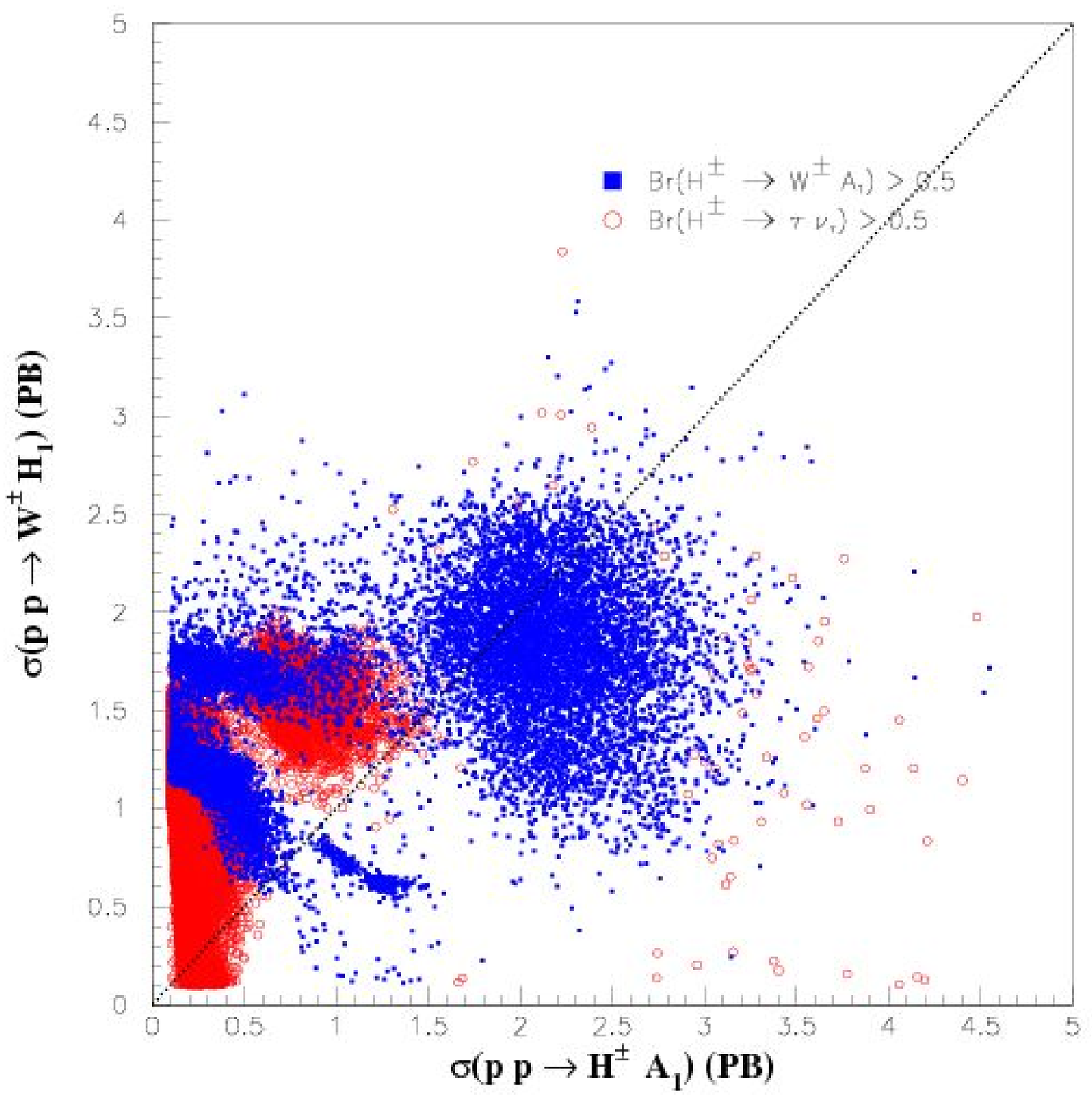}} & \hspace{-1.cm}
\resizebox{88mm}{!}{\includegraphics{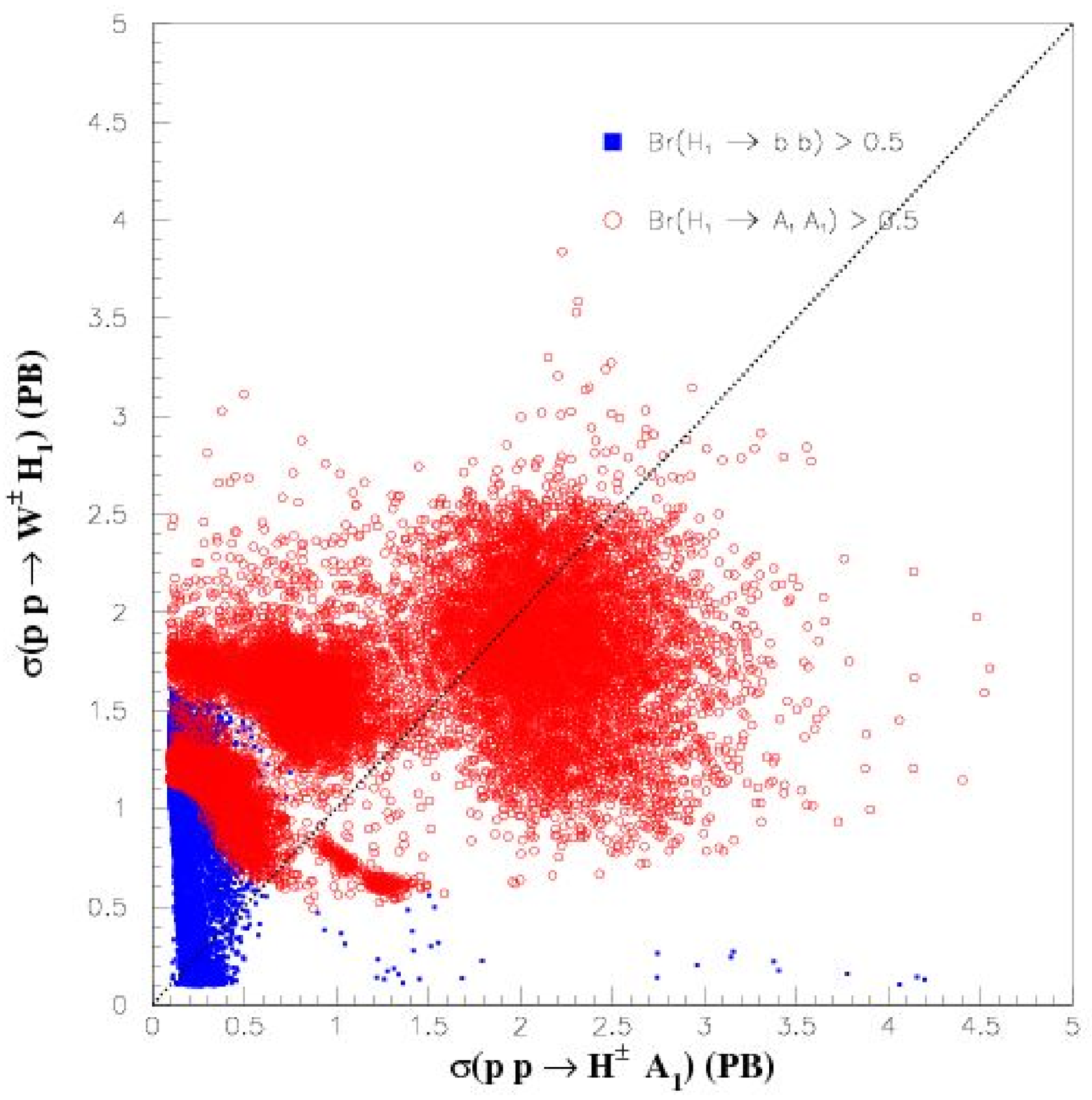}}
\end{tabular}
\caption{\it Left panel: comparison of $\sigma(pp \rightarrow H^\pm A_1)$ 
and $\sigma(pp \rightarrow W^\pm h_1) $ with two $H^\pm$ decay 
modes.  Right panel: comparison of $\sigma(pp \rightarrow H^\pm A_1)$ 
and $\sigma(pp \rightarrow W^\pm h_1) $ with two $h_1$ decay 
modes. The dotted line corresponds to
$\sigma(pp \rightarrow W^\pm h_1) = \sigma(pp \rightarrow H^\pm A_1)$. }
\label{fig3}
\end{figure}

As emphasized earlier, the processes $pp \rightarrow H^\pm A_1$ and
$pp \rightarrow V h_1$ could lead to the same final state, $Wbbbb$ 
or $W\tau\tau\tau\tau$. Hence a numerical comparison of their cross sections
is of particular interest and is shown in Fig.~(\ref{fig3}), where
all points satisfy the following conditions:
\begin{eqnarray}
\sigma(pp \rightarrow H^\pm A_1) > 0.1 \ pb\, \qquad \rm{and} \qquad
\sigma(pp \rightarrow W^\pm h_1) > 0.1  \ pb \,.
\label{cond-vvh-ca}
\end{eqnarray}
Superimposed on Fig.~(\ref{fig3}a) and Fig.~(\ref{fig3}b) are the main decay 
modes of the charged Higgs boson and the decay neutral Higgs boson $H_1$ 
respectively. We further impose the following conditions:
\begin{eqnarray}
Br(H^\pm \to W^\pm A_1) > 0.5 \qquad \rm{and} \qquad
Br(h_1 \to A_1 A_1) > 0.5 \,,
\label{conditions}
\end{eqnarray}
and the surviving points are displayed in Fig. (\ref{fig4}a). Importantly, 
there are many points where the two cross sections are of comparable size. 
We note that for these points in Fig. (\ref{fig4}a)  the pseudoscalar $A_1$ 
can be both R-axion like or a mixture of the three allowed basic axions.
If the magnitude of the cross sections of both $pp \rightarrow H^\pm A_1$ and 
$pp \rightarrow V h_1$ are similar then the interference of the two channels 
(i.e., the same $Wbbbb$ signature arising from distinct production mechanisms)
should be taken into account. We have neglected such effects in the present 
study.

We now discuss whether the $Wbbbb$ signatures can be distinguished 
experimentally by comparing the strategies adopted in 
\cite{Ghosh:2004wr} (for $pp\to H^\pm A^0$)
and \cite{Cheung:2007sv} (for $pp\to W^\pm h_1$).
In order to reconstruct the peak of the CP-even Higgs $h_1$,
one can select events with a charged lepton and four tagged $b$ quark jets 
as shown in \cite{Cheung:2007sv}.
This enables both a clean Higgs signal 
with high significance and a measurement of
$M_{h_1}$ given by the invariant mass of the four $b$ quark jets,
$m_{4b}$. The process $pp\to H^\pm A_1$ might be an irreducible
background but presumably could be significantly 
suppressed with the aforementioned cut on $m_{4b}$
e.g., $m_{h_1}-15 {\rm GeV} < m_{4b} < m_{h_1}+15 {\rm GeV}$.

Regarding detection of $pp\to H^\pm A^0$, it was demonstrated 
in~\cite{Ghosh:2004wr}
(for the analogous process $pp\to H^\pm H_1\to W H_1 H_1$ in the 
CP violating MSSM) that the mass of $H^\pm$ can be reconstructed.
This is achieved by defining a tranverse mass ($M_T$) 
which is a function of the momenta of the two secondary $b$ jets 
(i.e., those originating from the decay $H^\pm\to A_1W\to Wbb$) 
and the momenta of the lepton and missing energy 
coming from the $W$ boson. It was shown that
$M_T$ is sensitive to the underlying charged Higgs mass and 
thus can be used for the determination of $M_{H^\pm}$.
The pair of $b$ jets from $pp\to W^\pm h_1$ might be an irreducible
background but presumably could be suppressed with a cut on $M_T$

To reconstruct the peak of the light CP-odd neutral
Higgs $A_1$ one can require events with
four tagged $b$ jets, construct the three possible double pairings of 
$b\bar{b}$ invariant masses, and then select the pairing giving the least
difference between the two $b\bar{b}$ invariant masses values
\cite{Ghosh:2004wr}.
$W 4b$ signatures from the process $pp\to W^\pm h_1$ also 
contribute constructively to the reconstruction of $A_1$.
Thus we conclude that it is promising to reconstruct the peaks of 
the CP-even neutral Higgs ($h_1$), charged Higgs ($H^\pm$) 
and CP-odd neutral Higgs ($A_1$) and thus 
experimentally distinguish the $Wbbbb$ signatures arising 
from the two distinct production mechanisms. 
We defer a detailed simulation to a future work.

\begin{figure}
\begin{tabular}{cc}
\resizebox{88mm}{!}{\includegraphics{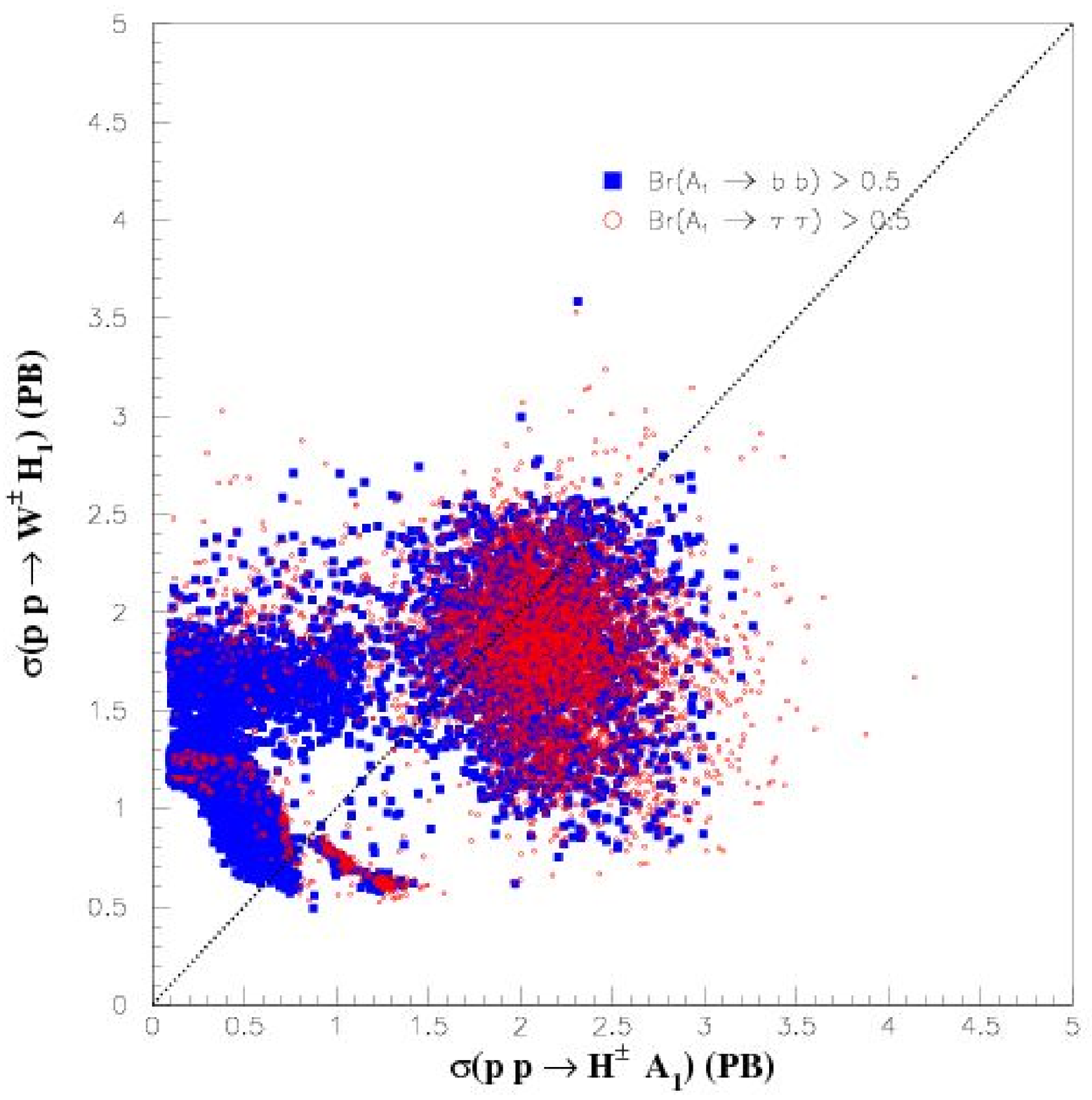}} & \hspace{-1.cm}
\resizebox{88mm}{!}{\includegraphics{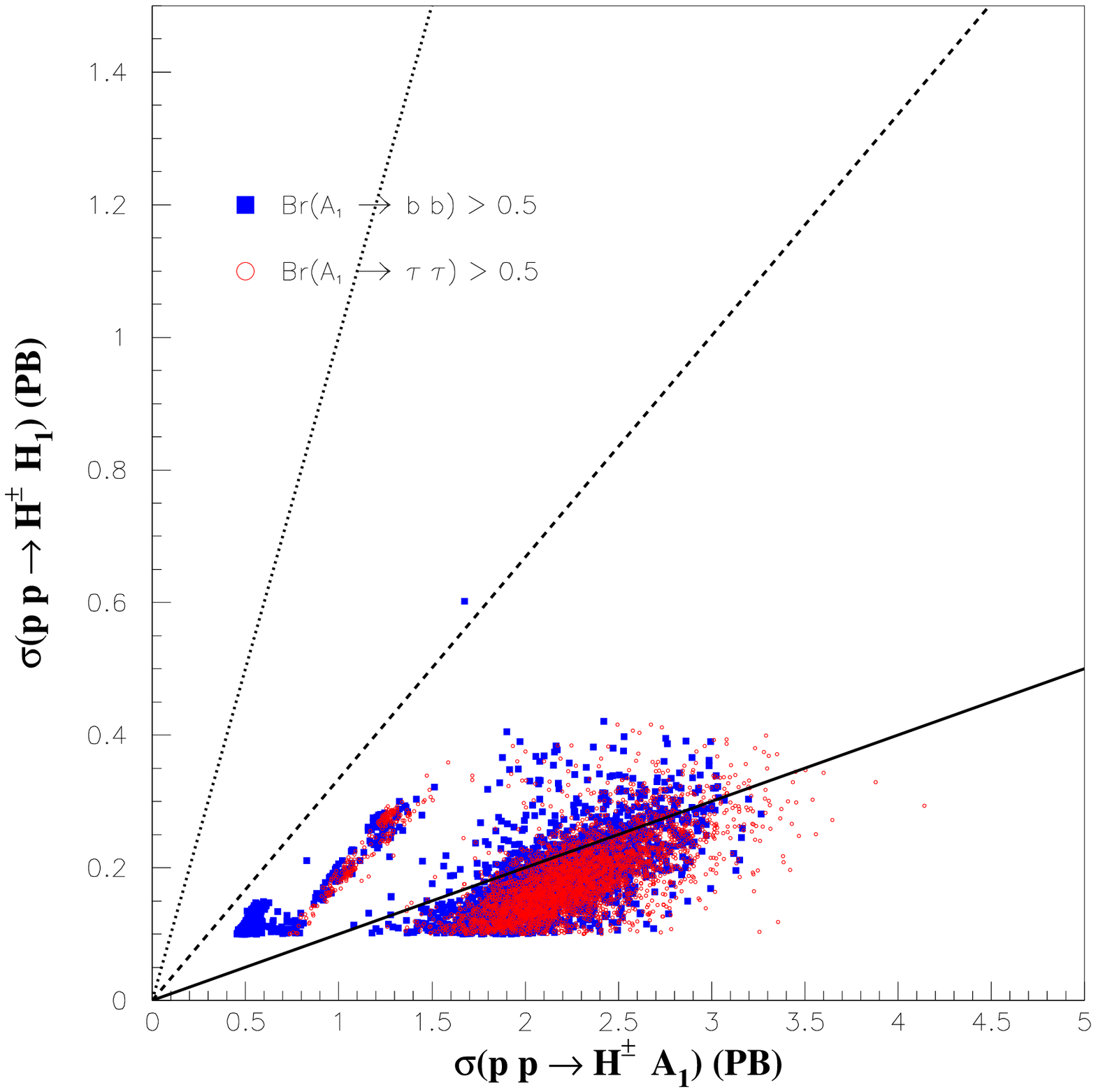}}
\end{tabular}
\caption{\it Left panel: comparison of $\sigma(pp \rightarrow H^\pm A_1)$ 
and $\sigma(pp \rightarrow W^\pm h_1) $ with different 
$A_1$ decay modes. Points are selected with the condition given in 
Eqs. (\ref{cond-vvh-ca}-\ref{conditions}). Right panel: comparison 
of  $\sigma(pp \rightarrow H^\pm A_1)$ and  $\sigma(pp \rightarrow H^\pm
h_1)$ with the same set of points.  
The dotted line corresponds to $\sigma(pp \rightarrow H^\pm A_1) = 
\sigma(pp \rightarrow H^\pm h_1)$; the dashed line corresponds to 
$\sigma(pp \rightarrow H^\pm A_1) = 3 \sigma(pp \rightarrow H^\pm h_1)$; 
the solid line corresponds to $\sigma(pp \rightarrow H^\pm A_1) = 
10 \sigma(pp \rightarrow H^\pm h_1)$. }
\label{fig4}
\end{figure}

Finally, we also compare the cross sections of $pp \rightarrow H^\pm A_1$ and
$pp \rightarrow H^\pm h_1$ in Fig.~(\ref{fig4}b). The points are 
from the same data sample used in Fig.~(\ref{fig4}a). 
It is clear that $\sigma (pp \rightarrow H^\pm A_1 )$ 
is around one order of magnitude 
larger than $\sigma(pp \rightarrow H^\pm h_1)$, 
and the underlying reason is that $M_{h_1} > 2 M_{A_1}$. 
Consequently, $pp\to H^\pm h_{1}\to W^\pm A_1h_{1}\to Wbbbb$ 
will also be suppressed and can be safely neglected.
Another interesting feature from Fig.~(\ref{fig4}b)
is that points satisfying the conditions listed 
in Eq. (\ref{conditions}) lead to $A_1$ composed
mainly of the doublet fields.
The conditions in Eq. (\ref{conditions}) together with the dominance of 
$h_1$ by doublet component (small $S_{13}$) 
can give large cross sections for both channels.
 
\section{Conclusion}
In summary, we have studied the phenomenology of light charged 
Higgs bosons in the framework of NMSSM. 
We performed a comprehensive study of the magnitude of the
branching ratios for the decays
$H^\pm\to W^\pm A_1$ and $H^\pm\to W^\pm h_1$ (first
considered in \cite{Drees:1998pw}).
It was shown that such decays can dominate over the standard
decays $H^\pm\to \tau^\pm\nu$ and $H^\pm\to tb$ 
both below and above the top-bottom threshold. 
This is due to the fact that $A_1$ can have
a large doublet component and small mass. 
Large branching ratios for 
$H^\pm\to W^\pm A_1$ and $H^\pm\to W^\pm h_1$ would affect
the anticipated search potential for $H^\pm$ at the LHC.

We also studied the production process
$pp\to H^\pm A_1$ and showed that sizeable cross sections
($> 1$ pb) are possible. We compared
the magnitude of the cross sections for both $pp\to H^\pm A_1$
and the Higgsstrahlung process $pp\to W^\pm h_1$
and showed that they can be of similar size. 
If $H^\pm$ and $h_1$ decay via $H^\pm\to W^\pm A_1$ and $h_1\to A_1 A_1$
respectively, the above two processes would lead to the same final 
state, $Wbbbb$ or
$W\tau\tau\tau\tau$. We stressed that the interference term for $Wbbbb$
and $W\tau\tau\tau\tau$ might not be negligible and should be taken
into account in any simulation study. In particular, 
the signature $Wbbbb$ affords promising detection prospects at the LHC 
and we discussed how to distinguish the distinct
contributions from $pp\to H^\pm A_1$ and  $pp\to W^\pm h_1$
by using appropriate cuts. 

It is known that intermediate values of $\tan\beta$ 
(e.g., $5 < \tan\beta < 20$) are
most problematic for discovery of $H^\pm$ at the LHC
since the $H^\pm tb$ Yukawa coupling (which is employed in the 
conventional production processes) takes its lowest values.
In such a region the process $pp \to H^\pm A_1$
can have a sizeable cross section if $m_{H^\pm}+m_{A_1} < 200$ GeV.
Therefore we propose $pp \to H^\pm A_1$ as a unique mechanism
to probe the parameter space of intermediate $\tan\beta$ and light charged 
Higgs boson in the NMSSM.

\section*{Acknowledgments} 
We would like to thank Kingman Cheung for useful discussions.
A.A is supported by the National Science Council of
R.O.C. under Grant No. NSC96-2811-M-008-020.
Q.S.Y is supported by the National Science Council of
R.O.C. under Grant No. NSC 95-2112-M-007-001
and 96-2628-M-007-002-MY3. 
A.G.A is supported by National Cheng Kung 
University Grant No. OUA 95-3-2-057.

\end{document}